\documentclass[twocolumn,superscriptaddress,floatfix,sort&compress]{revtex4-1}

\pdfoutput=1
\usepackage{graphicx}
\usepackage{amsmath}
\usepackage{color}
\usepackage{xr}
\begin{document}

\title{Shaping the Growth Behaviour of Biofilms Initiated from Bacterial Aggregates}

\author{Gavin Melaugh}
\affiliation{School of Physics and Astronomy, University of Edinburgh, James Clerk Maxwell Building, Peter Guthrie Tait Road, Edinburgh, EH9 3FD}
\author{Jaime Hutchison}
\affiliation{Center for Nonlinear Dynamics and Department of Physics, The University of Texas at Austin, Austin, Texas 78712-1199}
\author{Kasper N{\o}rskov Kragh}
\affiliation{Department of International Health, Immunology and Microbiology, Faculty Of Health Sciences, University of Copenhagen, DK-2200 Copenhagen}
\author{Yasuhiko Irie}
\affiliation{School of Life Sciences, Centre for Biomolecular Sciences, University of Nottingham, University Park, Nottingham NG7 2RD}
\affiliation{Department of Biology \& Biochemistry, University of Bath, Claverton Down, Bath BA2 7AY}
\author{Aled Roberts}
\affiliation{School of Life Sciences, Centre for Biomolecular Sciences, University of Nottingham, University Park, Nottingham NG7 2RD}
\author{Thomas Bjarnsholt}
\affiliation{Department of International Health, Immunology and Microbiology, Faculty Of Health Sciences, University of Copenhagen, DK-2200 Copenhagen}
\affiliation{Department for Clinical Microbiology, University of Copenhagen, DK-2100 Copenhagen}
\author{Stephen P. Diggle }
\affiliation{School of Life Sciences, Centre for Biomolecular Sciences, University of Nottingham, University Park, Nottingham NG7 2RD}
\author{Vernita D. Gordon}
\affiliation{Center for Nonlinear Dynamics and Department of Physics, The University of Texas at Austin, Austin, Texas 78712-1199}
\author{Rosalind J. Allen}
\affiliation{School of Physics and Astronomy, University of Edinburgh, James Clerk Maxwell Building, Peter Guthrie Tait Road, Edinburgh, EH9 3FD}

\begin{abstract}
Bacterial biofilms are usually assumed to originate from individual cells deposited on a surface. However, many biofilm-forming bacteria tend to aggregate in the planktonic phase so that it is possible that many natural and infectious biofilms originate wholly or partially from pre-formed cell aggregates. Here, we use agent-based computer simulations to investigate the role of pre-formed aggregates in biofilm development. Focusing on the initial shape the aggregate forms on the surface, we find that the degree of spreading of an aggregate on a surface can play an important role in determining its eventual fate during biofilm development. Specifically, initially spread aggregates perform better when competition with surrounding unaggregated bacterial cells is low, while initially rounded aggregates perform better when competition with surrounding unaggregated cells is high. These contrasting outcomes are governed by a trade-off between aggregate surface area and height. Our results provide new insight into biofilm formation and development, and reveal new factors that may be at play in the social evolution of biofilm communities.
\end{abstract}

\maketitle

\section*{Introduction}
\begin{figure}[t!]
\begin{center}
\includegraphics[scale=0.35] {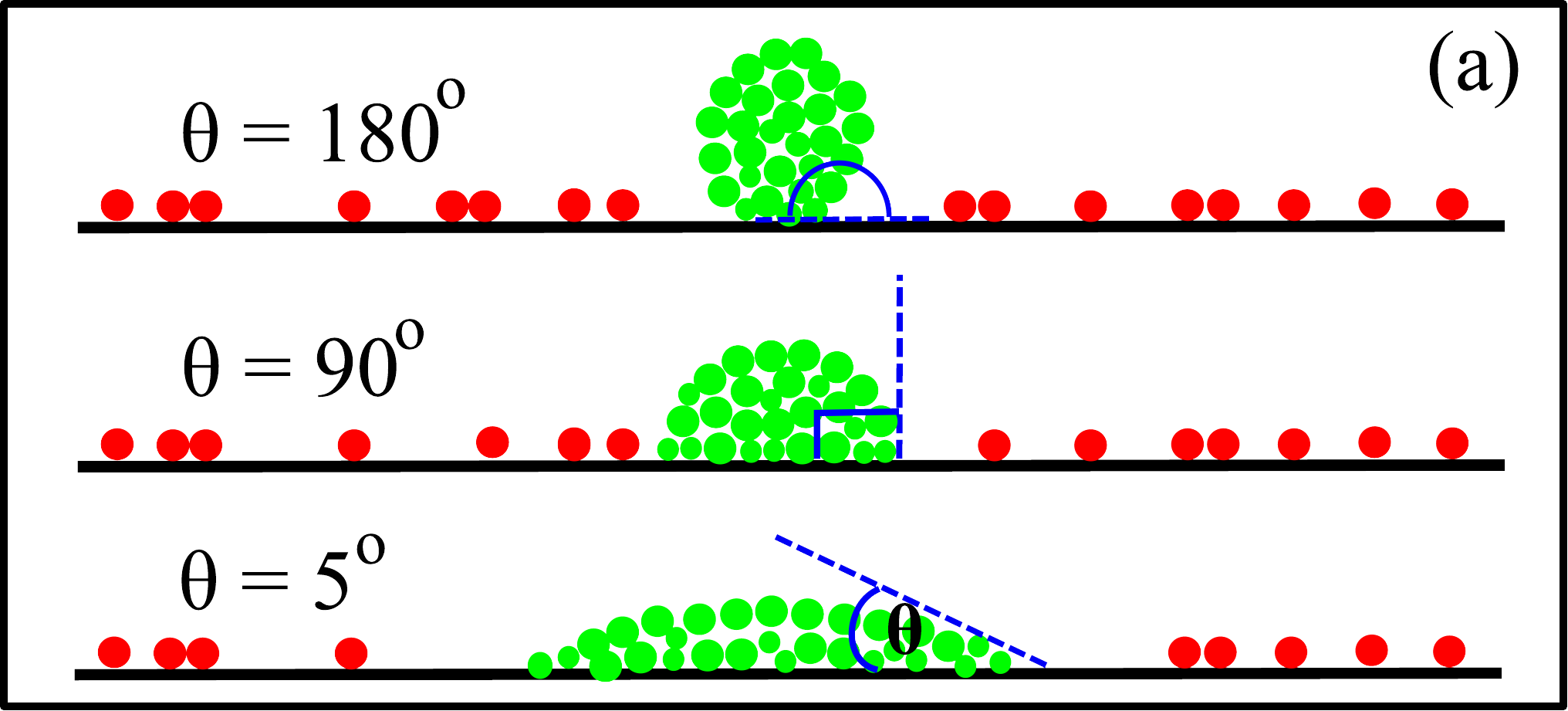}\\
\vspace{0.4cm}
\includegraphics[scale=0.65] {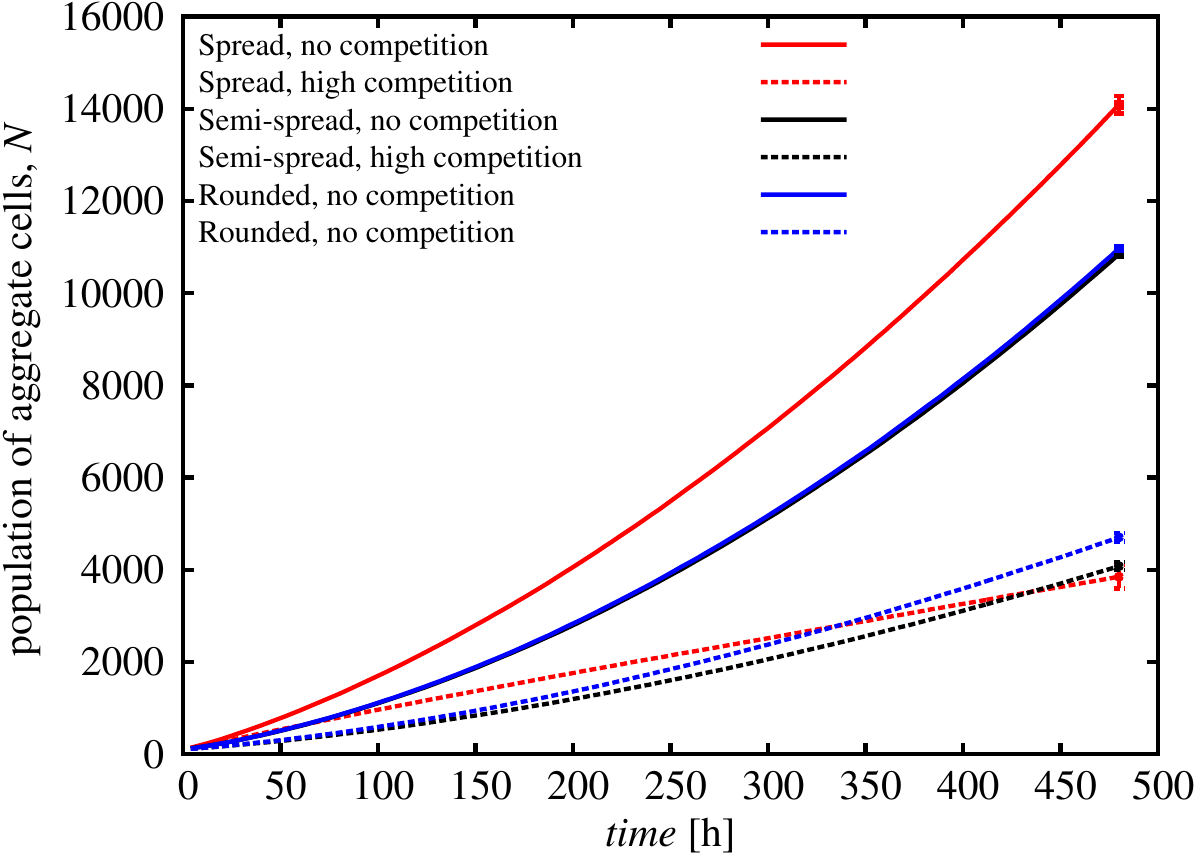} 
\caption[]{Our simulation set-up. (a) Schematic representation of bacterial aggregates (green) which are initially spread on a surface to varying extents. The schematic also shows surrounding, competing, unaggregated cells (red). Top- Rounded aggregate, $\theta=180^o$; Middle- Semi-spread aggregate, $\theta=90^{\circ}$; Bottom- Spread aggregate with $\theta \to 0^{\circ}$. Note that the size of the aggregates (in terms of number of bacteria) is approximately equal. (b)  Growth of the spread ($\theta=5^o$), semi-spread ($\theta=90^{\circ}$), and rounded aggregate ($\theta=180^{\circ}$) populations over the course of our simulations in the absence ($\rho=0$ cell $\mu$m$^{-1}$) and presence of competition ($\rho=0.5$ cell $\mu$m$^{-1}$). For clarity the error bars, representing the standard deviations, are only shown for the final data points. The standard deviations at these points are maximal.}
\label{fig:schematic}
\end{center}
\end{figure}

Surface-attached communities known as biofilms are believed to be the predominant mode of existence
for bacteria in many environmental settings \cite{Costerton1995}. Understanding how biofilms establish and grow is  also clinically important given their ubiquity in medical implant infections \cite{Costerton2005}, chronic wounds \cite{Fazli2009}, and in the respiratory tracts of cystic fibrosis patients \cite{Bjarnsholt2009a}. In the clinical context, biofilm communities often show enhanced virulence \cite{Burmølle2010}, resistance to antibiotics \cite{Hoiby2010}, and resistance to the host immune system \cite{Kharazmi1991}. These features may be associated with the spatial structure of the biofilm, which not only affects material transport, e.g., penetration of nutrients/antibiotics, but is also associated with differences in metabolism and gene expression among cells within the community \cite{Stewart2008, Sauer2002}.

In the canonical picture of biofilm development, individual cells land on a surface, attach and proliferate to form first micro-colonies and later 3-dimensional structures \cite{Monds2009}. However, bacteria are also known to form dense aggregated clumps when they are grown in liquid (planktonic phase) \cite{Bossier1996,Alhede2011, Haaber2012}. Moreover, cells often disperse from existing biofilms as clumps of aggregated cells.  Thus it is very likely that when a biofilm forms, some cells may arrive on the surface already in an aggregated state. In support of this view, evidence exists for the seeding of infections by pathogenic bacteria already in an aggregated state \cite{Faruque2006, Hall-Stoodley2005}, and bacterial aggregates are abundant in cystic fibrosis \cite{Bjarnsholt2009a, Burmølle2010} and tuberculosis \cite{Anton1996a} infections.

Having arrived on the surface, e.g., a plant leaf \cite{Monier2004}, a surgical implant \cite{Costerton2005}  or an industrial component \cite{Lens2003}, it is to be expected that cells within a bacterial aggregate will have to compete during biofilm development, both with other aggregates and with initially non-aggregated cells, to which they may or may not be genetically related. 

We take a first step towards understanding the role of pre-formed aggregates in biofilm development by investigating this competitive process, using agent-based simulations. Such simulations, in which the spatial structure of a biofilm emerges from local interactions between individual cells, have become a staple tool for investigating biofilm structure and dynamics \cite{Kreft2001,Xavier2005a,Alpkvist2006}, as well as social evolutionary aspects of biofilm development  \cite{Kreft2004, Nadell2010}. Using this approach, we  determine how a pre-existing aggregate of bacteria impacts the spatial structure of a biofilm, both in the presence and absence of competing unaggregated bacterial cells. 

Our main focus here is on the role of the initial shape of the aggregate. It is well known that bacterial interactions with a surface depend on features such as extra-cellular polymeric substances (EPS), presence of cell surface appendages (such as pili), and cell surface charge, which are species- and strain-dependent \cite{Tuson2013}. Moreover, soft-matter science has established that the nature of material-surface interactions can drastically affect the shape of fluid or semi-fluid droplets on surfaces \cite{Israelachvili2011}. It is therefore reasonable to suppose that in some  circumstances, bacterial aggregates will spread out in contact with  a surface, while in other scenarios, aggregates will adopt a more compact configuration. Here we investigate the biological consequences of aggregate shape in the seeding of biofilm growth. 

Simulating the development of biofilms initiated from initially spread or rounded aggregates, we find that the initial configuration of a bacterial aggregate on a surface is crucial in determining its eventual fate within the biofilm. In the absence of competitor cells on the surface, aggregates that maximise the extent to which they initially spread on the surface perform better than rounded ones because their initial access to nutrients (in the surrounding media) is greater.  However when faced with strong competition from neighbouring unaggregated cells, initially rounded aggregates perform better at long times, despite the fact that the rounded aggregate shape has a smaller surface area and hence a reduced exposure to nutrients. Importantly, we show that in an initially rounded aggregate, cells at the top of the aggregate proliferate at the expense of cells in the aggregate centre. This has interesting possible consequences for social evolution given that cooperation within clumps of aggregated cells has been suggested to be a stepping stone in the evolution of multicellularity \cite{Biernaskie2015,West18082015}

Our study highlights the effects of nutrient gradients and bacterial aggregate shape on long-term biofilm development. Our work reveals that these factors alone can produce a trade-off between nutrient access and competition, with the balance between these factors depending sensitively on aggregate shape. While the link between biofilm spatial structure and nutrient access has been highlighted in many other studies \cite{Stewart2008,Nadell2010,Hermanowicz2001,Stewart2003,Klapper2002}, our work is the first to focus on the role of preformed aggregates in this context. Our study should help to decipher the role of pre-formed aggregates in biofilm infections. More generally, our findings emphasise the need to consider pre-formed aggregates in our current understanding of biofilm development.

\section*{Methods}

In this study, we use agent-based computer simulations to model the growth of a biofilm on a surface, starting from initial configurations of bacterial cells like those shown in Figure \ref{fig:schematic}(a). In our simulations, an initial aggregate of cells (shown in green in Figure \ref{fig:schematic}(a)), adopting a particular shape, seeds an inert surface, and may compete with surrounding unaggregated cells (red in Figure \ref{fig:schematic}(a)).  Note that the red and green bacterial cells differ only in the manner in which they are initially arranged on the surface. At the start of our simulations, the ``red" bacteria are distributed at random across those parts of the surface not occupied by the aggregate (see Sections S1 to S5). To vary the extent of competition between the aggregated and unaggregated cells, we varied the initial cell density (number of cells per unit length of surface) of the unaggregated ``red" cells (see Sections S1 to S5). As a control, we also ran simulations in which the aggregate grew in the absence of the unaggregated cells.

The focus of this work is on the shape of the initial cell aggregate. Figure \ref{fig:schematic}(a) illustrates three different scenarios, in which the cell aggregate adopts a compact rounded shape (top), spreads out on the surface (bottom), or adopts an intermediate shape (middle). In each scenario, the dimensions of the aggregate are adjusted so that the number of cells within the aggregate remains the same (see Sections S1 and S2).

\subsection*{Aggregate Shape Characteristation}
To characterise quantitatively the aggregate shapes in the different scenarios shown in Figure \ref{fig:schematic}(a), we define the ``aggregate-surface angle" $\theta$, which is the angle that the initial aggregate makes with the flat surface. A small value of $\theta$ ($\theta \to 0^{\circ}$) describes an initial aggregate configuration that is spread on the surface, whereas a large value of  $\theta$ ($\theta \to 180^{\circ}$) describes an aggregate that is rounded. Given that the total number of cells in the initial aggregate is fixed, $\theta$ also encapsulates an interplay between the initial surface coverage of the aggregate and its initial height; with increasing $\theta$, the surface coverage decreases whereas height increases. In soft matter science, an analogous parameter is often used to describe wetting interactions between liquid droplets and surfaces \cite{Israelachvili2011}. A similar approach has recently been applied to the surface-spreading behaviour of eukaryotic cell aggregates \cite{Douezan2011}. From a phenotypic perspective, $\theta$ is related to the nature of the interactions between cells in the aggregate and between cells and the surface, and thus could be tunable by biological regulatory processes, or by evolution. We performed simulations for a range of $\theta$ values between 5$^{\circ}$ and 180$^{\circ}$ (Figure \ref{fig:schematic}(a)).

The aggregate configurations that we used to initiate our simulations were generated by ``transplantation" of circular segments from simulation snapshots of pre-grown biofilms (see Sections S1 and S2). This procedure proved preferable to other initialisation methods as it ensures no overlap between individual bacteria and enables the generation of different aggregate shapes of the same number density ($\sim$100 cells per unit area). By varying the radius of the circular segment we are able to ensure that each aggregate contained $\sim$100 cells that were initially spread on the surface to different extents. To ensure statistical accuracy, four different configurations were generated for each aggregate shape, defined by its value of $\theta$, and for each of these configurations, five simulations were performed using different random number seeds. Changing the random number seed affects the order in which individual bacteria grow and divide, and also changes the locations of the unaggregated cells on the surface surrounding the aggregate. A total of twenty simulations were therefore performed for each value of $\theta$, enabling us to sample both variation in the configuration and the ordering of cell updates (Section S5). Increasing the number of repeated simulations did not affect our results.

In common with many other biofilm simulation studies \cite{Kreft2001,Kreft2004,Nadell2010,Xavier2007}, our simulations were performed in two dimensions for the purposes of computational efficiency.  We have verified, however, that our key findings are reproduced when we use 3D simulations (see Sections S3 and S10).

\subsection*{Simulation Implementation}
We use the agent-based microbial simulation package iDynoMiCs \cite{Lardon2011} to model biofilm growth, starting from configurations such as those shown in Figure \ref{fig:no_comp_confs}. In these simulations, individual bacterial cells are represented as spherical agents, which grow and proliferate conditional on the local nutrient concentration, and ``shove" each other apart to relieve local stresses within the biofilm. The order in which cells are selected to grow and divide is random during each global time-step of the simulation, as is the direction of cell division. In our simulations, the initial distribution of surrounding competitor cells on the surface is also random. The simulations use a spatial grid to track the local nutrient concentration field. Nutrient is assumed to diffuse towards the biofilm from above, with the concentration being fixed to a bulk value in a layer far from the biofilm. Within the biofilm itself, nutrient diffusion is hindered relative to the region outside the biofilm. Nutrient consumption by the bacterial cells leads to local gradients, which can have a strong impact on the structural features of the growing biofilm \cite{Stewart2008,Nadell2010,Hermanowicz2001,Stewart2003,Klapper2002}.  Periodic boundary conditions are imposed on both the nutrient concentration field and the particle coordinates in the horizontal direction.

From a mathematical perspective, nutrient is represented as a concentration field,  the dynamics of which are governed by the the reaction-diffusion equation
\begin{equation}
\frac {\partial S(\bf x)}{\partial t} =   \nabla \cdot (D_S({\bf x}) \cdot \nabla S({\bf x})) + r_S({\bf x}), 
\label{eq:reac:diff}
\end{equation}
where $S(\bf x)$ is the space (${\bf x}$)-dependent nutrient concentration, $D_S({\bf x})$ is the diffusion coefficient of the nutrient, and $ r_S({\bf x})$ is the consumption rate of the nutrient by the bacteria. 
The rate of nutrient consumption, $r_S({\bf x})$, is related to the growth rate of the bacteria, ${\rm d}X/{\rm d}t$, via
\begin{equation}
r_S({\bf x}) = \frac{{\rm d}S}{{\rm d}t} = -\frac{1}{Y_{x/s}} \hspace{0.1cm}  \frac{{\rm d}X}{{\rm d}t}, 
\end{equation}
where $X({\bf x})$ is the local biomass density, and $Y_{x/s}$ is a yield coefficient that describes the amount of nutrient required to produce one unit of biomass $X$. 

The growth rate of each cell is governed by the well-known Monod function 
\begin{equation}
\frac{{\rm d}X}{{\rm d}t} = \mu_{max} \hspace{0.1cm} \frac{S}{k_S + S} X, 
\label{eq:monod}
\end{equation}
where $\mu_{max}$ is the maximum specific growth rate of the bacteria, and $k_S$ is the concentration of nutrient, $S$, at which the growth rate is half maximal. The growth parameters used in our simulations were taken from empirical and simulation studies on {\it Pseudomonas aeruginosa}, assuming glucose to be the rate-limiting nutrient (see Table \ref{table:gr_parameters}). However our results are not sensitive to the detailed parameter choice, or to the choice of rate-limiting nutrient.  Note that the growth rate parameters $Y_{X/N}$,  $\mu_{max}$, and $k_S$ are the same for both the aggregate cells and the competitor cells (see Table \ref{table:gr_parameters}).

From a practical point of view, in idynoMiCs the nutrient concentration fields are assumed to be in pseudo steady-state with respect to biomass growth and therefore the time dependence is removed from Equation \ref{eq:reac:diff}
\begin{equation}
0 =   \nabla \cdot (D_S({\bf x}) \cdot \nabla S({\bf x})) + r_S({\bf x}). 
\end{equation}
\begin{table*}[t]
 \begin{center}
   \begin{tabular}{l l l l}
    \hline\hline
    Symbol \ & Description \  & Value \ & Notes/ref \\ [0.5ex]
  \hline
$[S]_{bulk}$        & Bulk concentration of limiting nutrient &  $5.4 \times 10^{-3}$ gL$^{-1}$ & Within range of values from \cite{Xavier2005, Xavier2007, Lardon2011} \\
$Y_{X/N}$    & Yield coefficient for Monod equation (Equation \ref{eq:monod}) &   0.44 & Within range of values from \cite{Characklis, Bailey1986, Xavier2007}  \\
$\mu_{max}$ & Maximum specific growth rate & 0.35 h$^{-1}$ & Within range of values from \cite{Characklis, Bakke1984, Robinson1984,Beyenal2003} \\
$K_N$    & Half saturation concentration of nutrient & $3 \times 10^{-3}$    & Within range of values from \cite{Characklis, Bailey1986, Horn2001, Alpkvist2006, Lardon2011, Xavier2005a, Bakke1984, Robinson1984}\\
$\gamma$       & Density of biomass &  200 gL$^{-1}$ & \cite{Characklis, Xavier2005a, Xavier2007} \\
$D_G$       & Diffusivity of glucose in water &  $5.8 \time 10^{-5}$ m$^2$day$^{-1}$ & \cite{Shedlovsky1955} \\ 
$L_x$       & Dimension of system in horizontal direction &  1032 $\mu$m & Ensures aggregates do not interact periodically\\ 
$L_y$       & Dimension of system in vertical direction &  1032 $\mu$m & Corresponds to the horizontal length\\ [0.1ex]
$L_{dbl}$       & Thickness of diffusion boundary layer &  80 $\mu$m & Within range of values from \cite{Picioreanu1998, Alpkvist2006, Xavier2005a}\\ [0.1ex]
 \hline\hline
   \end{tabular}
\caption[]{Input parameters for biofilm simulations.}
\label{table:gr_parameters}
 \end{center}
 \end{table*}
This equation is solved numerically for every global update of the bacterial population.
In our simulations, we used a bulk nutrient concentration of $5.4 \times 10 ^{-3}$ gL$^{-1}$ (see Section S6) comparable with previous work \cite{Lardon2011,Xavier2007,Xavier2005}. Each of our simulations was run for a total time of 480 h, in order to explore the long-term growth dynamics (see results and Section S7).

Our complete parameter set is listed in Table \ref{table:gr_parameters}. Using the parameters in Table \ref{table:gr_parameters}, our simulations produce spatially structured biofilms 200-300 $\mu$m in height after a simulation time of 480 h (see Figure \ref{fig:comp_no_comp_grs}, and Section S6). 


\section*{Results}
\subsection*{Initial Aggregate Shape Determines Growth Dynamics}
\label{subsection:Initial Aggregate Shape Determines Growth Dynamics}
To assess the growth dynamics of pre-formed aggregates in a biofilm, we tracked the number of progeny cells, $N$, produced by aggregates of different shape (Figure \ref{fig:schematic}(b)). 
We investigated three different aggregate shapes, characterised by the angle $\theta$ (See Methods). The three angles, $\theta = 5^{\circ}, 90^{\circ}, 180^{\circ}$, describe pre-formed aggregates that are initially arranged on the surface in either a spread, intermediately spread, or rounded manner.  To investigate the effect of the surrounding unaggregated competitor cells (red cells in Figure \ref{fig:schematic}(a)), we varied the density, $\rho$, of these cells between two extremes regimes of competition reported in this study: $\rho = 0$ cell $\mu$m$^{-1}$, no competition; and $\rho = 0.5$ cell $\mu$m$^{-1}$, high competition. 

Not surprisingly, aggregates grow better in the absence of surrounding cells on the surface regardless of their initial shape, i.e., for $\rho =0$ cell $\mu$m$^{-1}$. In this ``non-competitive regime", the initially spread aggregate produces more progeny than the more rounded aggregate. This is evident in Figure \ref{fig:no_comp_confs}, which shows representative initial aggregate configurations and the structures of the biofilms which they form after 480 hours. 

Competition from unaggregated competitor cells on the surfaces leads to more complex behaviour. Figure \ref{fig:schematic}(b) shows that, in the presence of strong competition, the spread aggregate produces more progeny over short times than the rounded aggregate. However, over longer times, the size of the population arising from the rounded aggregate is larger. 

For an aggregate in the presence of competing non-aggregating cells, Figure \ref{fig:schematic}(b) points to two strategies for maximising progeny. For long-lived biofilms, progeny can be maximised  by the aggregate adopting a rounded configuration, whereas if the biofilm is short-lived then it may instead be optimal for the aggregate to spread on the surface. 

\subsection*{Initial Aggregate Shape Affects Long-Term Biofilm Structure}
\begin{figure*}[t!]
\begin{center}
 \includegraphics[scale=0.3]{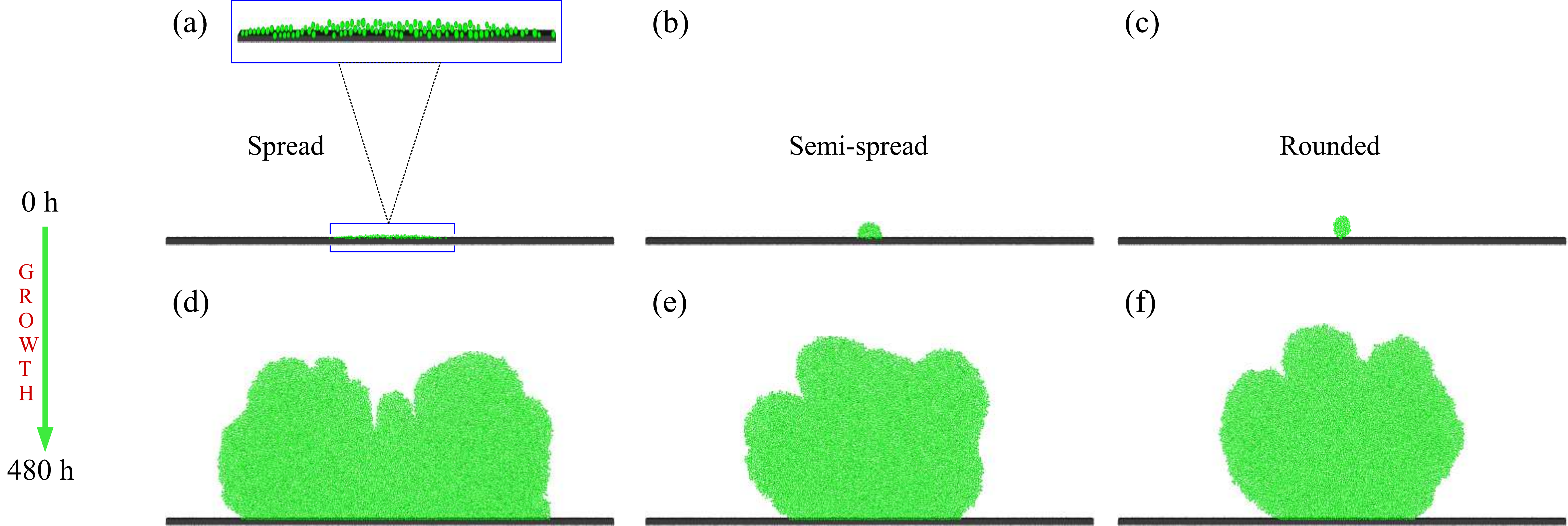} \\
\caption[]{Initial aggregate arrangement affects biofilm morphology. Simulation snapshots of three bacterial aggregates initially arranged on the surface and the biofilms they form after 480 h: (a) Spread, 0 h. A zoomed in image is also shown to make the shape of the aggregate easier to resolve; (b) Semi-spread, 0 h; (c) Rounded, 0 h; (d) Spread, 480 h; (e) Semi-spread, 480 h; (f) Rounded, 480 h.}
\label{fig:no_comp_confs}
\end{center}
\end{figure*}

In our simulations the initial shape of the aggregate influences the long-term structure of the biofilm.
Figure \ref{fig:no_comp_confs} shows typical biofilm structures formed after 480 h of growth, starting from aggregates that were initially spread on the surface ($\theta = 5^{\circ}$, left), rounded ($\theta = 180^{\circ}$, right) or partially spread ($\theta = 90^{\circ}$, centre), in the absence of competition from surrounding cells. It is clear that the spread aggregate covers much more of the surface during growth than its more rounded counterparts.

\begin{figure}[t!]
\begin{center}
 \includegraphics[scale=0.25]{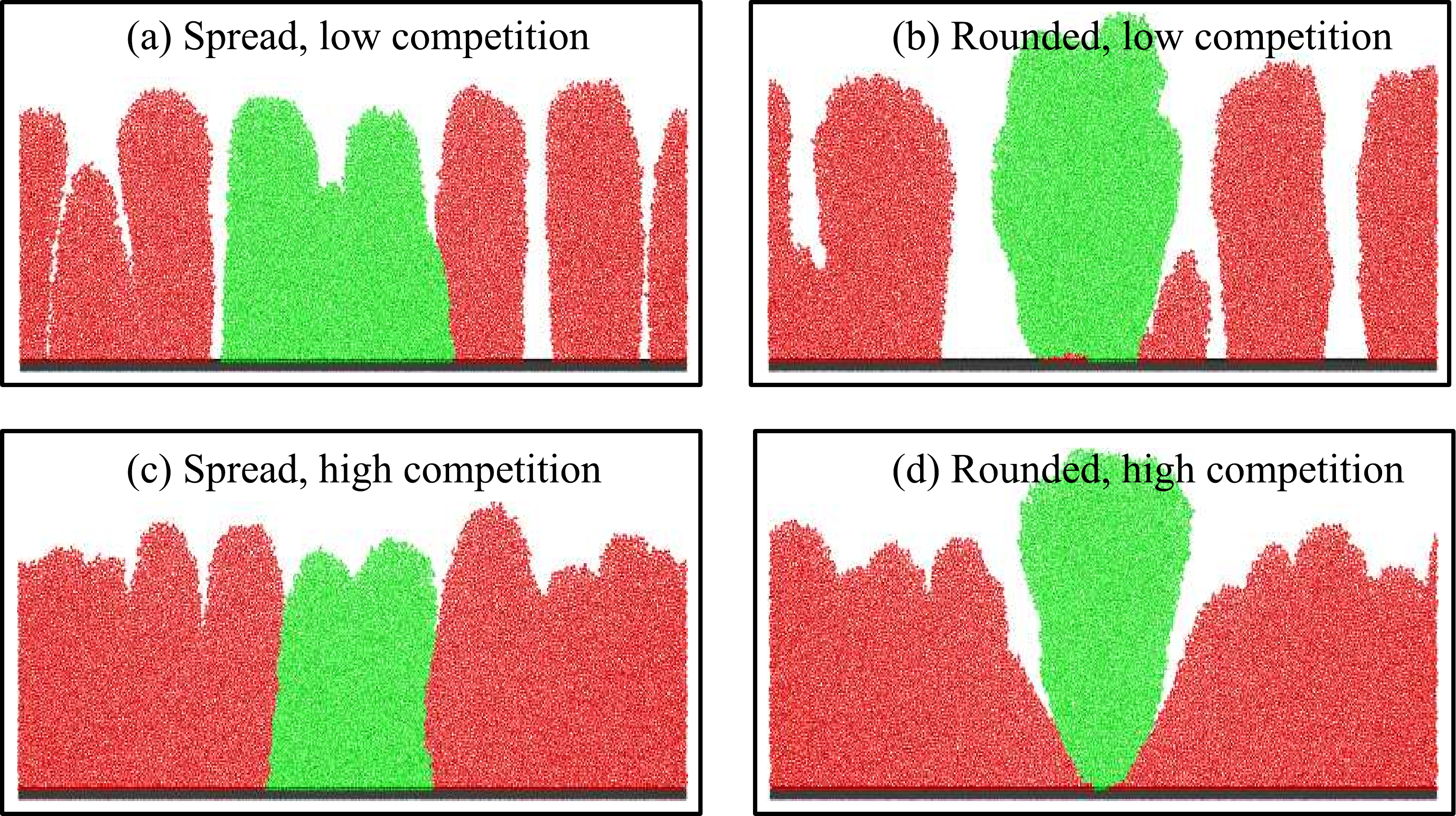} \\
\caption[]{Aggregate shape and neighbouring strain density affect biofilm morphology. Simulation snapshots of biofilms seeded from spread and rounded aggregates after 480 h growth in the presence of a low and high density inoculum of the competing strain: (a) $\theta = 5^{\circ}, \rho = 0.01$ cell $\mu$m$^{-1}$; (b) $\theta = 180^{\circ}, \rho = 0.01$ cell $\mu$m$^{-1}$; (c) $\theta = 5^{\circ}, \rho = 0.5$ cell $\mu$m$^{-1}$; (d) $\theta = 180^{\circ}, \rho = 0.5$ cell $\mu$m$^{-1}$.}
\label{fig:comp_snapshots}
\end{center}
\end{figure}
In the presence of competition (Figure \ref{fig:comp_snapshots}), we observe a marked difference in the structure of the biofilms that originate from spread aggregates (left panels) and from rounded aggregates (right panels). For the spread aggregate (a and c), the green section of biofilm that originates from the aggregate is structurally indistinguishable to that of the surrounding red biofilm that originated from the competing, unaggregated cells. In contrast, for the rounded aggregate (b and d), cells originating from the aggregate form a distinct ``clump", which is taller than the surrounding biofilm. When the density of competing (red) cells is high (Figure \ref{fig:comp_snapshots}(d)), there is a cell-free gap around the growing clump that appears to be a result of nutrient depletion.

It is clear from the biofilm structures shown in Figures \ref{fig:no_comp_confs} and \ref{fig:comp_snapshots} that, even at very long times, the spatial structure of a biofilm can be affected by the initial spatial configuration of its founder cells (see also Section S6). While it might seem remarkable that apparently small changes in initial configuration can have dramatic effects on biofilm structure even after many cell generations, this effect is in fact well-known in a different context. For initially flat biofilms, Dockery and Klapper showed theoretically that small inhomogeneities in initial configuration may be magnified into large ``fingers" over the course of biofilm development \cite{Klapper2002}. This phenomenon is shown as a fingering instability and arises from the fact that an emerging protrusion (or finger) is elevated above, and thus depletes nutrients from the surrounding biofilm. This leads to positive feedback, in which the enhanced growth of the cells at the top of the instability is to the detriment of the surrounding cells below \cite{VanLoosdrecht2002}.

While Dockery and Klapper assumed that structural inhomogeneities would arise spontaneously during biofilm growth, in our simulations such inhomogeneities are effectively created by the presence of the initial aggregates.
The introduction of the rounded aggregate amongst the lawn of unaggregated cells on the surface at high competition leads to an instability in the biofilm structure that propagates as the biofilm develops.

\subsection*{Nutrient gradients are important determinants of aggregate fate}
\begin{figure}[h!]
\begin{center}
 \includegraphics[scale=0.12]{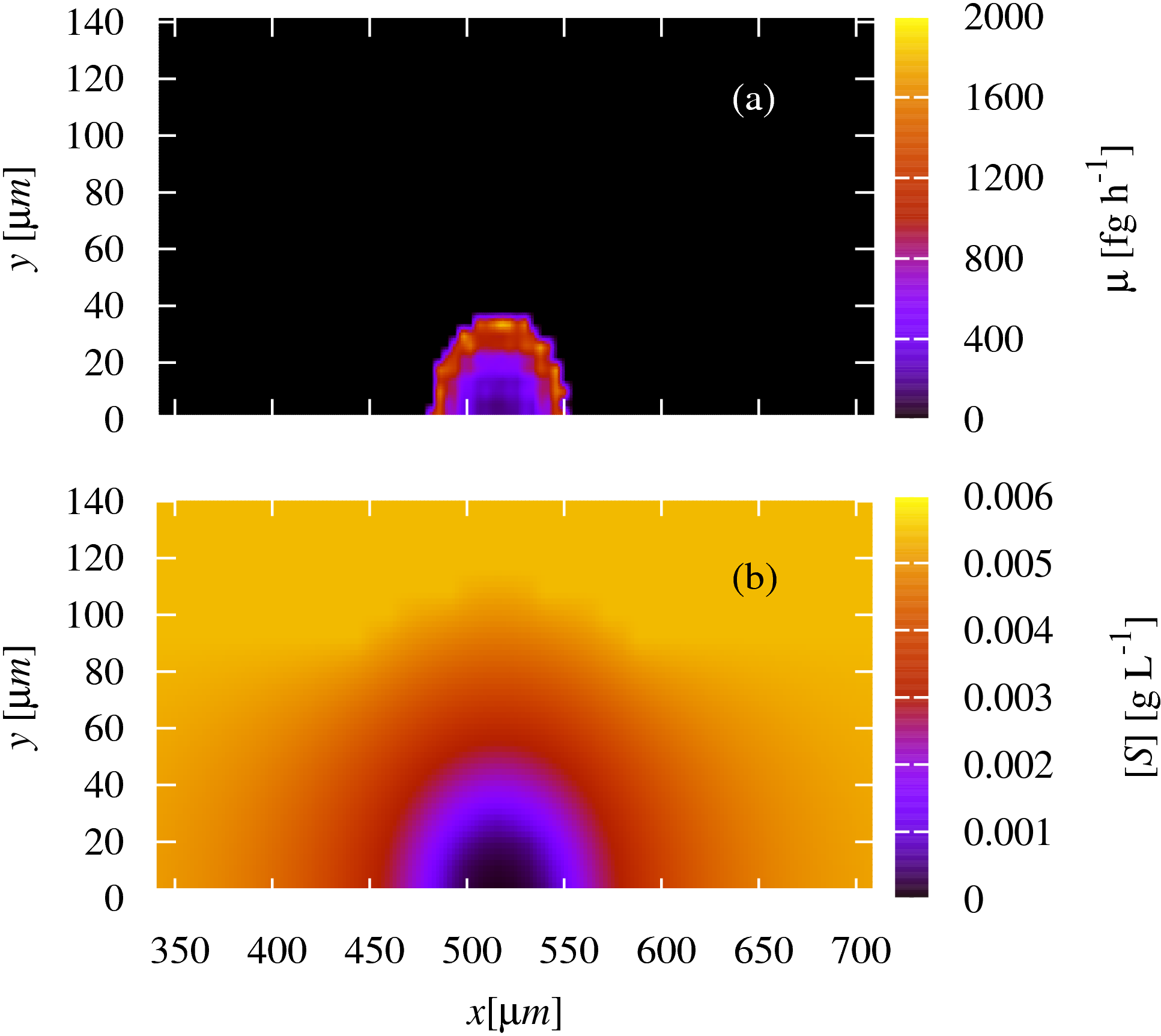}
\caption[]{Cells on the outside of the aggregates grow faster because they have greater access to nutrients. (a) Cell growth rate ($\mu$) distribution of the biofilm formed from the semi-spread aggregate in the absence of competition after 4h. (b) Corresponding nutrient concentration field, $[S]$.}
\label{fig:sol_gr}
\end{center}
\end{figure}

\begin{figure*}[t!]
\begin{center}
 \includegraphics[scale=0.16]{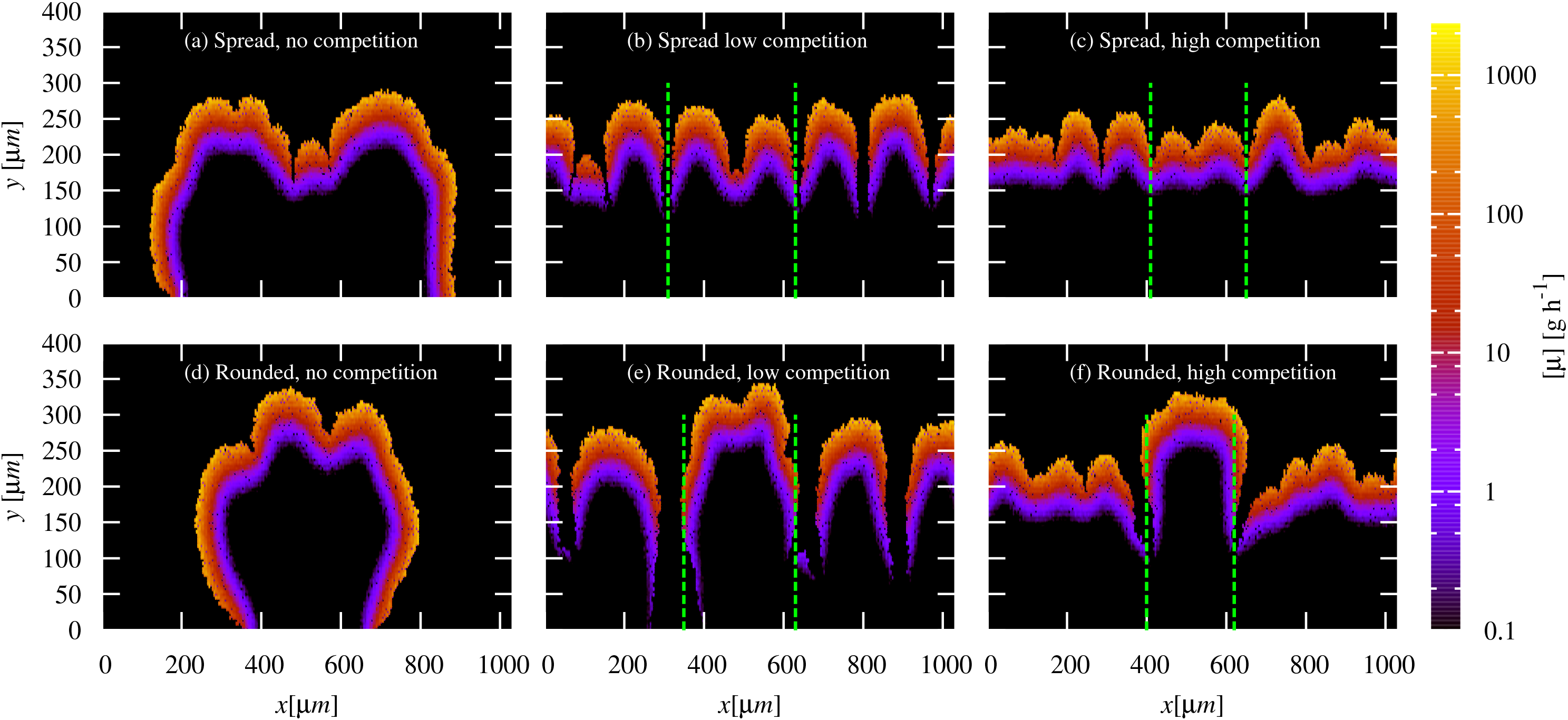} \\
\caption[]{Gradients in individual cell growth rates emerge in our simulated biofilms during growth. Cell growth rate distributions for the spread and rounded aggregates at after 480 h of growth: (a) $\theta = 5^{\circ}, \rho = 0.0$ cell $\mu$m$^{-1}$; (b) $\theta = 5^{\circ}, \rho = 0.01$ cell $\mu$m$^{-1}$; (c) $\theta = 5^{\circ}, \rho = 0.5$ cell $\mu$m$^{-1}$; (d) $\theta = 180^{\circ}, \rho = 0.0$ cell $\mu$m$^{-1}$; (e) $\theta = 180^{\circ}, \rho = 0.01$ cell $\mu$m$^{-1}$; (f) $\theta = 180^{\circ}, \rho = 0.5$ cell $\mu$m$^{-1}$. These distribution correspond to the configurations in Figures \ref{fig:no_comp_confs} and \ref{fig:comp_snapshots}. Note that the gradient in cell growth rate is so large that a log scale is used for visualisation purposes. The green dashed lines represents an approximate boundary between the aggregate cells and the surrounding competing strain.}
\label{fig:comp_no_comp_grs}
\end{center}
\end{figure*}

It is well known that growth rate heterogeneities, resulting from nutrient concentration gradients, emerge during biofilm growth \cite{Wentland1996}.
With this in mind, we tracked the growth rates of individual cells as a function of their position within the growing biofilm. 
Even in the very early stages of biofilm growth, we see heterogeneity in growth rates which emerge from (and influence) spatial gradients in nutrient concentration. Figure \ref{fig:sol_gr} illustrates this for a semi-spread aggregate ($\theta = 90^o$), after 4 h of growth, in the absence of competition. As expected, the cell growth rate is highly heterogeneous across the biofilm, Figure \ref{fig:sol_gr}(a), with cells on the outside growing faster than those on the inside because they have better access to nutrients (Figure \ref{fig:sol_gr}(b)).

The growth rate heterogeneities shown in Figure \ref{fig:sol_gr}(a) are amplified in the later stages of biofilm growth. Figure \ref{fig:comp_no_comp_grs} shows the spatial distribution of cell growth rates for biofilms arising from spread and rounded aggregates after 480 h, in the presence and absence of competitor cells, for the same simulations as shown in Figures \ref{fig:no_comp_confs} and \ref{fig:comp_snapshots}. 

In all cases we observe, as in previous work \cite{Nadell206}, two distinct regions of growth activity within the developing biofilms: an outer layer of metabolically active cells and an interior region of inactive cells. These distinct regions arise because consumption by cells in the outer layer deprives cells in the inner layer of nutrients \cite{Nadell206}. We also observe a large gradient in individual cell growth rate within the growing layer itself (note the logarithmic scale in Figure \ref{fig:comp_no_comp_grs}). The dynamics of the metabolically active layer determine the overall growth behaviour and structure of the biofilm. In Section S9 we show that the active layer of the rounded aggregate, unlike the spread aggregate, continues to expand in the presence of competition; explaining why its total population becomes larger than that of the spread at longer times in Figure \ref{fig:schematic}(b), and why it's structures tend to fan outwards at the top (Figure \ref{fig:comp_no_comp_grs}(f)).
 
Although it is well documented that nutrient gradients arising during biofilm growth play an essential role in biofilm formation \cite{Stewart2008,Stewart2003,Nadell2010}, so far few studies have investigated the effect of pre-formed bacterial clumps in this process; in particular how the initial arrangement of cells within a clump affects biofilm structure and development. Figures \ref{fig:sol_gr} and \ref{fig:comp_no_comp_grs} show that the initial arrangement of cells on the surface can determine the shape and structure of a growing biofilm because small initial differences in nutrient gradients become amplified as the biofilm develops. 

\subsection*{Competition for Nutrient Favours Rounded Aggregates}
\begin{figure}[t!]
\begin{center}
\includegraphics[scale=0.75] {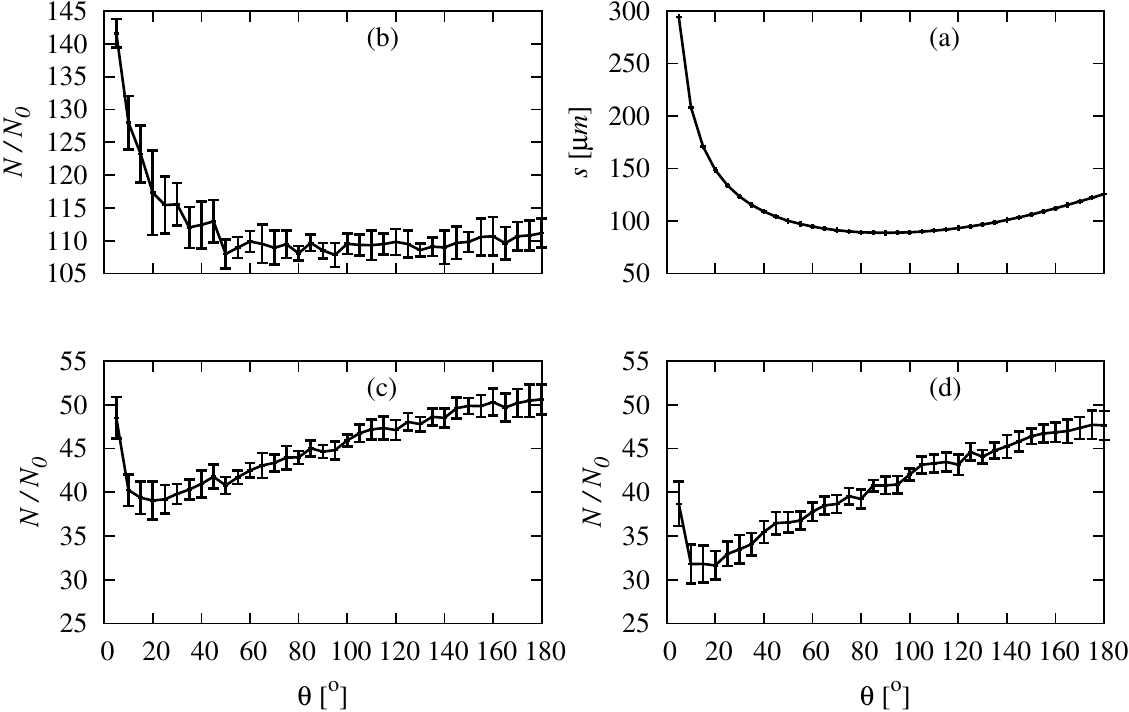}
\caption[]{Average number of progeny, $N/N_0$, of aggregates defined by their surface-aggregate angle $\theta$, the functional from of which changes with increasing density of competitor cells: (a) $\rho = 0$ $\mu$m cell$^{-1}$; (b) aggregate-medium interface length, s, as a function of $\theta$; (c) $\rho = 0.145$ $\mu$m cell$^{-1}$; $\rho = 0.5$ $\mu$m cell$^{-1}$. Vertical bars in represent the standard deviation from 20 data points.}
\label{fig:fit_curves}
\end{center}
\end{figure}

Next we investigated how the fate of an aggregate, as measured by the average number of progeny of one of its cells, varies with aggregate shape.
To this end, we computed the number of progeny cells, $N$, arising from the aggregate after a period of biofilm growth, relative to the initial number of cells in the aggregate, $N_0$, for a range of aggregate shapes, determined by $\theta$, at varying levels of competition. 
For this analysis, we carried out long simulations (480 h of biofilm development), so that the ratio $N/N_0$ reflects the long-time fate of the progeny of cells within the aggregate (see Section S7). 

Figure \ref{fig:fit_curves}(a) shows $N/N_0$ plotted as a function of the aggregate-surface angle, $\theta$, in the absence of competition from surrounding unaggregated cells. It is clear that the spread aggregate produces more progeny on average than the rounded one.

In the previous section we saw that cells on the outside of the aggregate have more access to nutrients even in the very early stages of biofilm growth. Thus, we might hypothesise that the growth advantage of the spread aggregate, in the absence of competition, is related to its larger surface area. Indeed, Figure \ref{fig:fit_curves}(b) shows that 
$N/N_0$ correlates closely with the interfacial area (or arc length) of the initial aggregate, s($\theta$). We therefore conclude that the spread aggregate produces more progeny than the rounded one because the former has a greater surface area with the surrounding medium, providing greater exposure to nutrient in the initial stages of growth.
The difference in initial structure between the spread and rounded aggregates therefore translates into significant differences in cell fate, even after many generations of biofilm growth. 

\begin{figure}[t!]
\begin{center}
\includegraphics[scale=0.4] {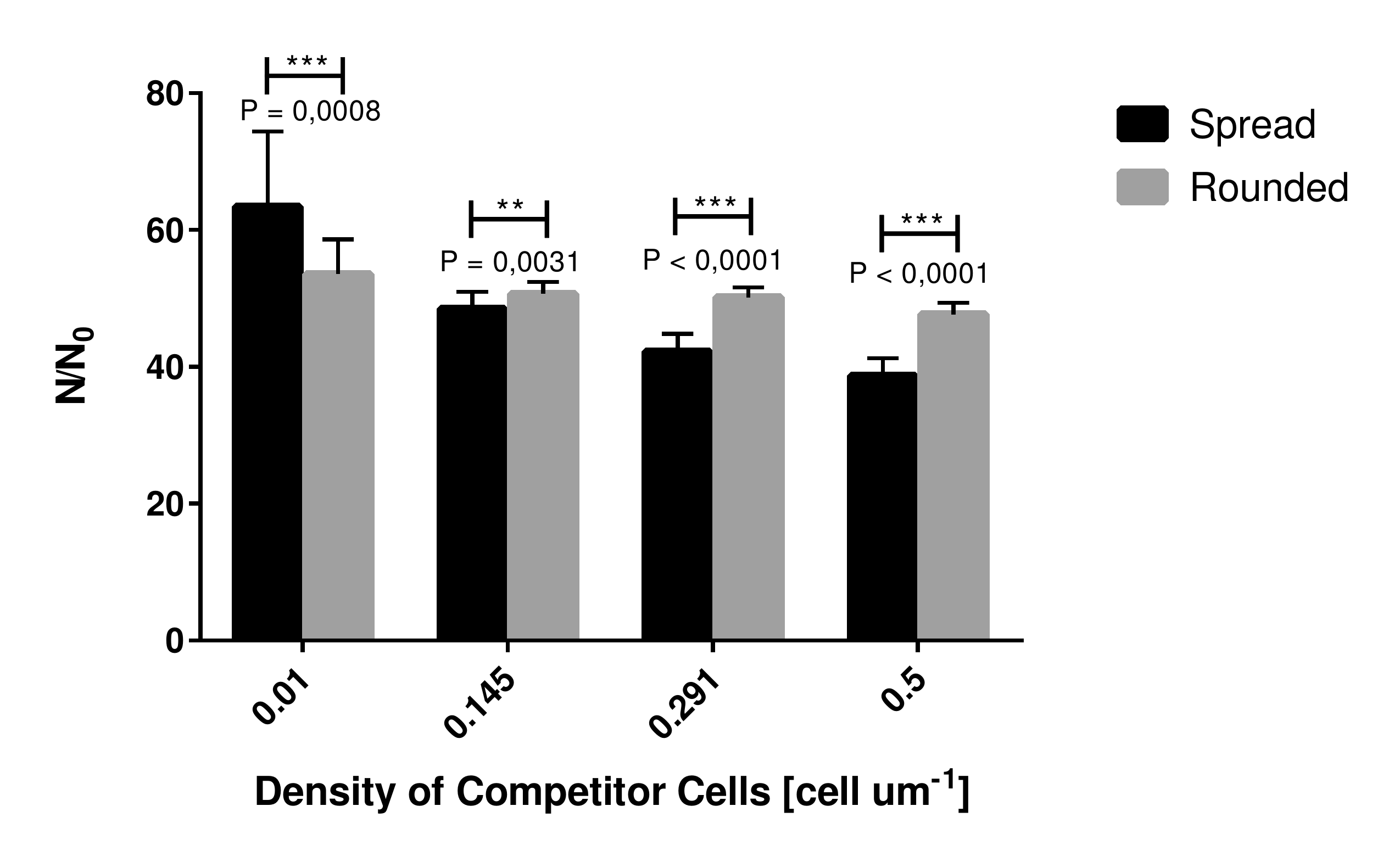}
\caption[]{Relative fitness as measured by $N/N_0$ of rounded aggregates increases with competition. Rounded aggregates become favourable relative to spread aggregates with increasing density of competitor cells.}
\label{fig:fit_bar}
\end{center}
\end{figure}

As the density of competition of unaggregated cells on the surface increases, however, a very different scenario emerges. Figures \ref{fig:fit_curves}(c) and (d) show that cells in the rounded aggregate (large $\theta$)  produce more progeny, on average, than those in the spread aggregate. This is more evident in Figure \ref{fig:fit_bar}, which shows $N/N_0$ for the spread and rounded aggregates as a function of the density of competitor cells. This effect is also observed at higher nutrient concentration (see Section S8).

Why does competition from surrounding unaggregated cells favour rounded over spread aggregates? Close inspection of Figure \ref{fig:fit_curves} shows that, while the number of progeny produced by the spread aggregate decreases with increasing competition from surrounding cells (panels (c), and (d)), the number of progeny produced by the rounded aggregate remains rather constant. 

\begin{figure*}[t!]
\begin{center}
\includegraphics[scale=0.15] {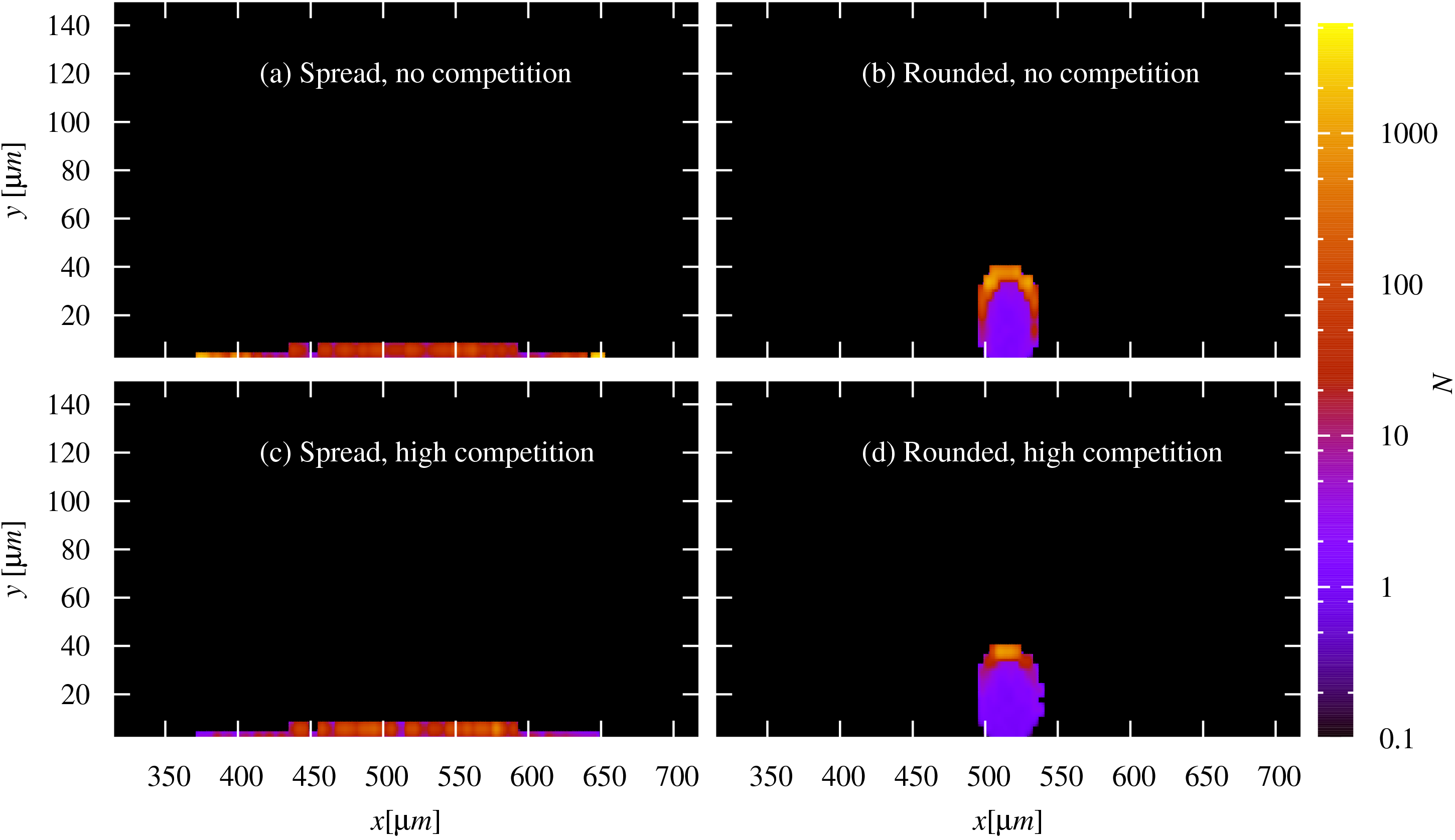}
\caption[]{Distribution of fittest cells varies with aggregate shape. 2D histograms representing the number of progeny, $N$, produced (480 h) by individual bacteria as a function of their initial location in the spread and rounded aggregates in the absence and presence of competition: (a) $\theta = 5^{\circ}, \rho = 0.0$ cell $\mu$m$^{-1}$; (b) $\theta = 180^{\circ}, \rho = 0.0$ cell $\mu$m$^{-1}$; (c) $\theta = 5^{\circ}, \rho = 0.5$ cell $\mu$m$^{-1}$; (d) $\theta = 180^{\circ}, \rho = 0.5$ cell $\mu$m$^{-1}$. Note that these distributions were averaged over 20 trajectories for each aggregate. Note that the gradient in the number of progeny is so large that a log scale is used for visualisation purposes.}
\label{fig:cell_fit}
\end{center}
\end{figure*}

This finding can be understood by investigating how the fate of an individual cell within an aggregate depends on its initial spatial location. To this end, we tracked the number of progeny of each individual founder bacterium, as a function of its initial position within an aggregate. This constitutes a local, spatially-resolved version of the ``fitness measure" $N/N_0$. Averaging our results over 20 repeated simulations allowed us to generate a map showing the average number of progeny produced by individual cells within an aggregate, for initially spread and rounded aggregates, Figure \ref{fig:cell_fit}.

Figure \ref{fig:cell_fit} shows that the initial position within an aggregate indeed has a strong effect on cell fate. In the absence of competition from surrounding unaggregated cells, the most successful cells in the spread aggregate are those at the horizontal extreme edges; in the interior region of the aggregate, cell fate is more uniform (Figure \ref{fig:cell_fit}(a)). It seems likely that in this case, cells at the horizontal edges have an advantage because their progeny can expand in the horizontal direction, whereas the progeny of cells in the interior of the aggregate must compete with their neighbours within the aggregate for nutrients and space. The proliferation of the cells at the edges of the aggregate drives the lateral expansion of the growing biofilm which we observe in Figures \ref{fig:no_comp_confs}(a) and (d). In contrast, for the rounded aggregate (Figure \ref{fig:cell_fit}(b)), cell fate is overall more heterogeneous within the aggregate, with the most successful cells being located around the outside surface of the aggregate. For the rounded aggregate, it appears that height is a relevant factor as well as the proximity to the aggregate surface. 

Figure \ref{fig:cell_fit}(c) shows that, in the presence of competition, cells at the horizontal edges of the spread aggregate actually do less well than those in the centre. The decreased fitness of these cells explains the inhibited lateral expansion observed in Figures \ref{fig:comp_snapshots}(a) and (c).
For the rounded aggregate, the most successful cells in the absence of competition are those at the top of the aggregate. In the presence of competition (Figure \ref{fig:cell_fit}(d)), these cells, which are now highly localised at the top, are elevated above the level of the competitor cells and therefore are little affected by the increased competition for nutrients. The ``fitness" cost associated with its smaller surface interface is compensated in the presence of competition by its height, since its top cells remain unchallenged by competitors with respect to nutrient access. 

\section*{Discussion}
Given the tendency of many bacteria to aggregate, and the frequent observation of aggregates in diverse environmental situations \cite{Burmølle2010,Monier2003,Stoodley2002a}, it seems likely that natural biofilms are often initiated from pre-formed aggregates. Despite this, the role of pre-formed aggregates in biofilm development has, to our knowledge, not yet been addressed. In this paper, we have investigated the fate of pre-aggregated cells during biofilm formation, using individual-based simulations. Our study shows that an initial aggregate can have a significant and long-lasting effect on biofilm spatial structure, even after many generations of cell growth. Focusing on the role of aggregate shape, we find that, in the absence of competition for nutrients from surrounding cells, an aggregate that is initially spread on the surface is favoured over one that is initially rounded even over long periods of biofilm development. This is likely to be because the spread aggregate has initially a larger surface over which it can absorb nutrients, giving it an initial growth advantage that is then maintained as the biofilm grows.

Strikingly though, our results change qualitatively in the presence of competition from surrounding, unaggregated cells. When this competition is strong, although the spread aggregates still grow faster in the early stages of biofilm development, rounded aggregates become more successful (produce more progeny) as the biofilm develops over longer times. This effect appears to arise from a trade-off between height (as nutrients diffuse from above) and exposed surface area. In the absence of competition, surface area is more important than height, and the spread aggregate is favoured. However, in the presence of competition, height becomes more important, since cells at the top of the aggregate can avoid competing for nutrients with the surrounding competitors. Since the rounded aggregate is taller than the spread aggregate, it gains a ``fitness" advantage under conditions of strong competition that is only realised after long times. 

Bacterial biofilm formation is a complex phenomenon which involves a plethora of biological mechanisms including cell motility \cite{Klausen2003,Klausen2003}, EPS production \cite{Flemming2007,Flemming2010}, metabolic and other phenotypic differentiation \cite{Sauer2002,Stoodley2002a,Watnick2000}, and cell-cell interactions such as quorum sensing \cite{Davies1998,Miller2001,Diggle2007}.  In our simulations, almost all of this biological complexity has been neglected; our model takes account only of nutrient gradients established by cell consumption, nutrient-dependent growth, and competition among cells for space. Nevertheless this simplistic approach produces biologically interesting, and potentially testable, predictions. In particular our simulations predict that being initially spread on a surface is a better strategy for a bacterial aggregate in the absence, but not in the presence, of competition. Understanding how further biological complexity might affect this picture would be a very interesting topic for further work. Another avenue worth investigating would be the effects of biofilm erosion and the subsequent detachment of cells. Here, our simulations have not included the effects of fluid flow, which among other effects, may flatten the biofilm by detaching protruding cells.
 
How might aggregates of different shape arise in nature? It is well known that bacterial interactions with surfaces can vary greatly depending both on the physical and chemical properties of the surface \cite{Dorobantu2009,Harimawan2011,Wright2010}, and on bacterial phenotypes such as EPS production and the presence of surface appendages. It is therefore very likely that aggregates formed from bacteria of different taxa or strains, landing on different surfaces, might adopt different configurations. For example, certain bacteria produce surfactant which can alter the morphology of a developing biofilm and allow them to expand over surfaces more efficiently \cite{Davey2003,Angelini27102009}.

Our work has been inspired by the observation that bacterial aggregates often form in the planktonic phase \cite{Bossier1996,Alhede2011, Haaber2012}. Aggregates are known also to form via the detachment of bacterial clumps from a mother biofilm \cite{Wilson2004,Hall-Stoodley2005,Stoodley2001a}; should such aggregates land on a pristine surface, similar phenomena to those discussed here would be expected to arise. Moreover, our results could also be relevant to aggregates that form on the surface itself. In the classical picture of {\it P. aeruginosa} biofilm development, individual cells land on a surface, upon which they migrate and proliferate to form small aggregates (i.e. microcolonies). Surface-induced motility mechanisms \cite{Harshey2003,Henrichsen1972} such as twitching \cite{Burrows2012}, crawling \cite{Conrad2011} and walking \cite{Gibiansky2010} have been implicated in this process. Once formed, such small aggregates would compete with surrounding cells for nutrients in much the same way as the pre-formed aggregates that we investigate in this paper. 

This study has focused on a single pre-formed aggregate seeding the surface and competing with initially unaggregated cells during biofilm formation. To further understand biofilms in nature, this work should be extended to investigate competition between multiple aggregates arranged on the surface, and competition between aggregates and mixed strains of bacteria, i.e., strains with different growth rates. 

Recently cooperation within clumps of aggregated cells has been suggested to be a major stepping stone in the evolution of multicellularity \cite{Biernaskie2015,West18082015}. Our study thus also hints that interesting social interactions might arise between cells within an aggregate. For all aggregate shapes, we observe heterogeneity in fitness among cells within the aggregate. This is particularly pronounced for the rounded aggregate, where cells at the top are strongly favoured while those in the centre of the aggregate hardly proliferate. Based on arguments recently put forward by West and Biernaskie \cite{Biernaskie2015,West18082015}, one might predict that rounded aggregates would be favourable under conditions where cells within the aggregate are closely related, whereas spread aggregates, in which fitness differences between cells are less pronounced, might form where cells are less closely related. This leads to interesting further questions, e.g., when a rounded aggregate initiates biofilm growth, does the majority of cells in the aggregate ``sacrifice" their future progeny in favour of their kin at the top? This idea supports previous suggestions that height plays a crucial role in competition within biofilms \cite{Xavier2007}. While previous work pointed to EPS production as a means to push progeny cells above the surrounding competitors \cite{Xavier2007,Kim2014}, our work shows that aggregate formation also provides a means to this end. Such a picture raises new questions about the evolutionary implications of bacterial aggregation.

\section*{Acknowledgements}
GM would like to thank Diarmuid Lloyd and Bartek Waclaw for helpful discussions, and Jan Ulrich Kreft, Robert Clegg, and Kieran Alden for technical advice throughout the duration of this study. All authors wish to thank The Human Frontiers Science Program for financial support under Grant Number RGY 0081/2012. In addition GM and RA would also like to thank the EPSRC EP/J007404. Also, KK and TB were supported by the Lundbeck Foundation. KK was supported by Oticon Fonden and Knud H\o jgaards Fond. RJA was supported by a Royal Society University Research Fellowship.

\section*{Author Contributions}
GM designed and performed the simulation work, and drafted the manuscript. RA designed and coordinated the study and helped draft the manuscript. The idea of investigating the growth of pre-formed aggregates on surfaces was conceived by TB, VG, RA, and SG. KK, JH, YI, and AR all aided in the design of the study and the biological interpretation of the results. All authors provided critical evaluation of the study, assisted in editing the manuscript, and provided approval for submission. 

\section*{Competing Interests}
The authors declare no competing interests.

\bibliographystyle{naturemag}
\bibliography{biofilm_aggregates} 
\end{document}



\maketitle
\section{Generating 2D bacterial aggregates} \label{sec:gen_aggregates}

To generate bacterial aggregates of the appropriate shape, circular segments were extracted from simulation configurations of pre-grown biofilms as illustrated in Figures \ref{fig:gen_confs1} and \ref{fig:gen_confs2}.
Figure \ref{fig:gen_confs1}(a) shows a circle (red) with radius $R$ and segmental region defined by the white area between the black horizontal line and the perimeter of the circle. The ``surface-aggregate angle", $\theta$, defines the angle between the black horizontal line (surface) and the tangent line of the circle at the point of contact with the surface. 

The area, $A$, of a circular segment of radius $R$ and with surface-aggregate angle of $\theta$, is given by \cite{circ_seg_ref}
\begin{equation}
A = \frac{1}{2} (2\theta - {\rm sin}2\theta) R^2.
\label{eq:seg_area}
\end{equation}
The corresponding arc length, $s$, of an aggregate defines the initial interface between the bacterial aggregate and the surrounding nutrient medium, and is computed according to
\begin{equation}
s= 2R\theta.
\end{equation}
To ensure our aggregates contained the same number of cells ($\sim$100), the area in Equation \ref{eq:seg_area} was fixed at 1257 $\mu$m$^2$.
 
The procedure for generating the starting aggregate configurations was performed as follows:
\begin{enumerate}
\item An {\it in silico} biofilm was grown for a simulation time of 7 days. This simulation was initiated with 200 cells randomly distributed on the surface and a nutrient concentration of 5.4$\times$10$^{-2}$ g L$^{-1}$ was used. A typical biofilm resulting from such a simulation was approximately 800 ${\rm \mu}$m in height with some finger-like projections such as those in Figure \ref{fig:gen_confs2}. 
\item A point, $P$, was defined in the middle of the biofilm (Figure \ref{fig:gen_confs1} and \ref{fig:gen_confs2}). The point $P$ is the reference point from which the origin, $O$, of a circle is computed. The point $P$ was selected such that the density of cells within any segmental area was relatively uniform. 
Bacterial aggregates of shapes $\theta = 5, 10, 15, 20, 25,.....,180^o$ with fixed area of 1257 $\mu$m$^2$ were then generated  as in Figure \ref{fig:gen_confs2}.
\item For each $\theta$, the radius of the corresponding circle was computed by solving  Equation \ref{eq:seg_area} for $R$. Using the radius, $R$, of this new circle, we computed the distance, $X$,  to move the origin of this new circle above or below the point $P$ according to 
\begin{equation}
X = R\hspace{0.1cm}{\rm sin}\alpha, 
\end{equation}
where 
\begin{equation}
\alpha = 90-\theta.
\end{equation}

\begin{figure}[t!]
\begin{center}
\includegraphics[scale=0.6]{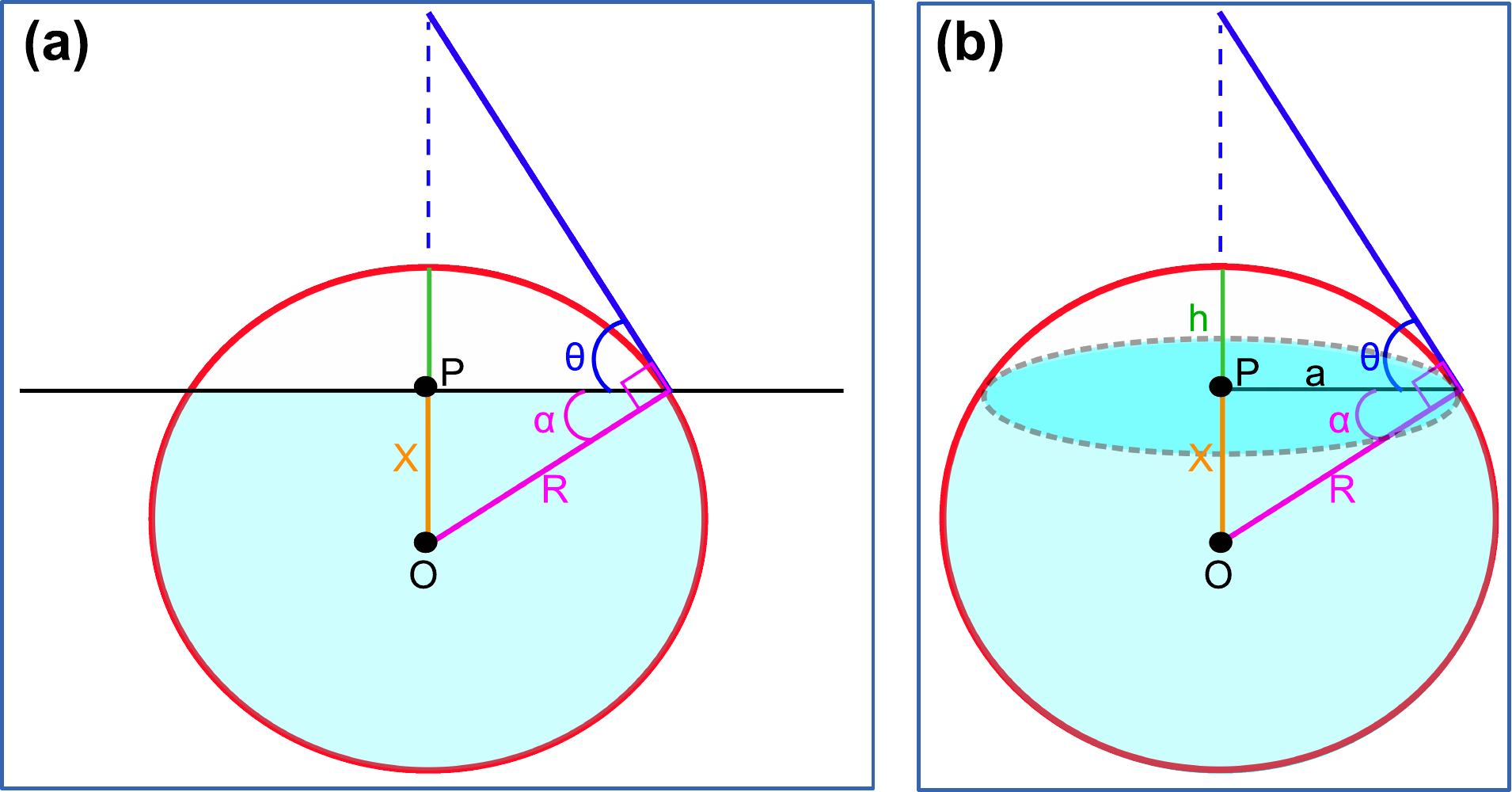}  \\
\caption[]{Schematic representation of the geometry of aggregates in (a) 2D and (b) 3D. The light blue region represents the area not included when computing cells within a distance $R$ from the origin, $O$. The angle $\theta$ is the contact angle of the aggregate with the surface.}
\label{fig:gen_confs1}
\end{center}
\end{figure}

\begin{figure}[t!]
\begin{center}
\includegraphics[scale=0.65]{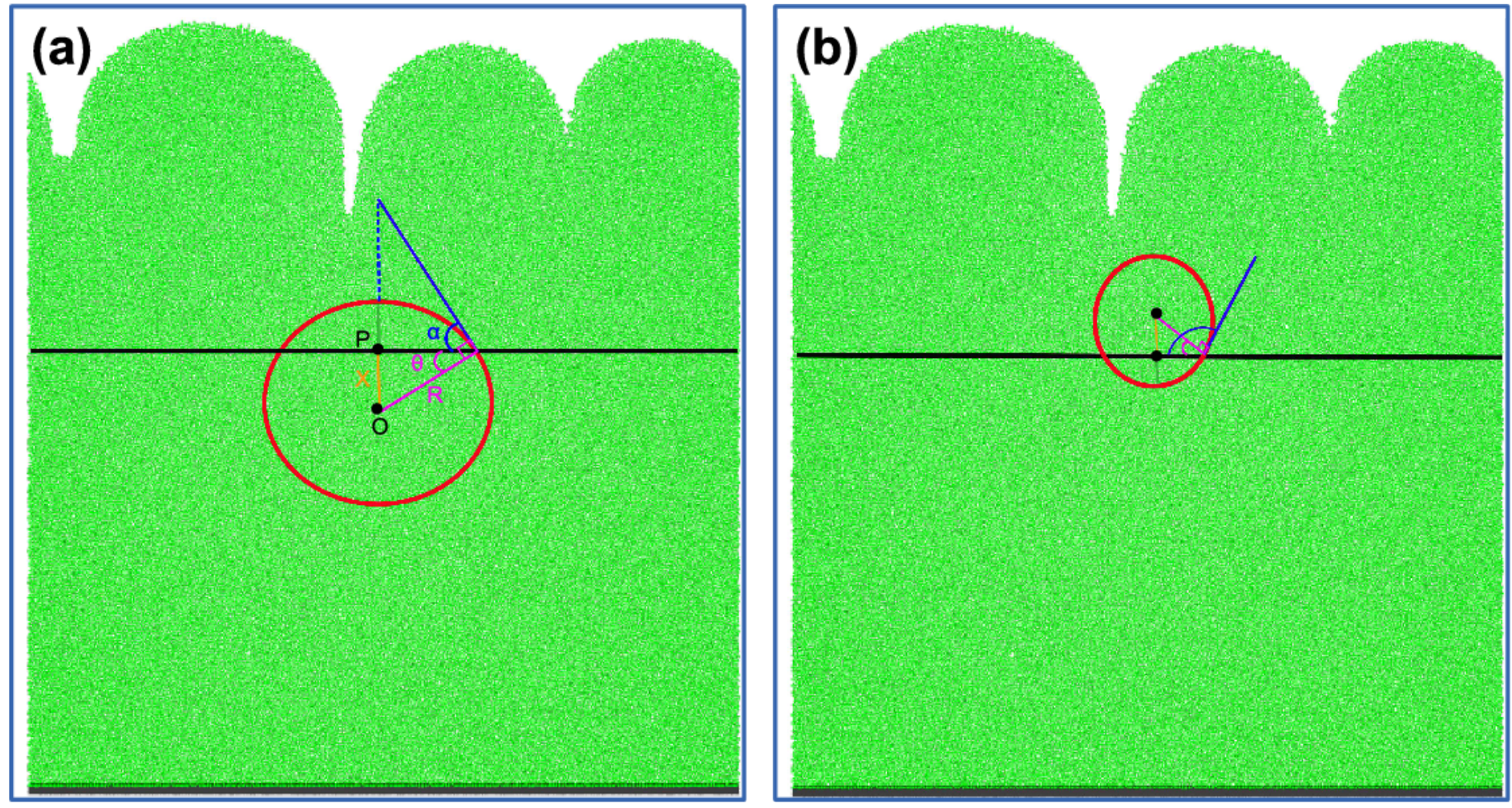}  \\
\caption[]{Generating aggregates with different geometries but containing the same number of cells. Using pre-grown 2D biofilm configurations, green, aggregates of the desired shape can be generated using the geometric relationships from Figure \ref{fig:gen_confs1}(a). (a) Schematic showing how a spread aggregate is generated. (b) Schematic showing how a more rounded aggregate is generated.}
\label{fig:gen_confs2}
\end{center}
\end{figure}

\item With the new origin and radius defined, we then extracted all bacteria within the radial distance from $O$. Note those bacteria located below the horizontal line (blue region) were excluded. 
The corresponding segment gave rise to a bacterial aggregate with a shape determined by the value of $\theta$.
\item The vertical component of the point $P$ was then subtracted from all bacterial coordinates so that the base of the aggregate was located on the surface as shown in Figure 1(a) of the main manuscript.
\end{enumerate}

\section{Generating configurations of competing ``single cells" in 2D} \label{sec:comp_single_cells}
The width of the surface available to the surrounding competitor cells depends on the surface coverage of the aggregate. As surface coverage of the aggregate increases, $\theta \to 0$, the amount of available surface either side of the aggregate decreases.  Surrounding cells were inserted at random positions on the surface excluding the region occupied by the aggregate. To ensure a constant density of cells surrounding the aggregate, the number of cells in these regions were varied according to the width of surface available.

\section{Generating 3D aggregates}\label{sec:gen_3D}
To generate aggregates in our 3D simulations, we performed an analogous process to that used for the 2D case. Configurations of spherical caps were transplanted from pre-grown biofilms using the geometrical relationships illustrated in Figure \ref{fig:gen_confs1}(b).
The volume of a spherical cap is  \cite{sph_cap_ref}
\begin{equation}
V_{cap}= \frac{1}{3} \pi R^3 (2 - 3{\rm sin}\alpha + {\rm sin}^3\alpha), 
\end{equation}
and the corresponding surface area of the cap is given by 
\begin{equation}
S = 2\pi R h,
\end{equation}
where the height, $h$, is given by 
\begin{equation}
h = R -X
\end{equation}
A spherical cap volume, $V_{cap}$, of 5575 $\mu$m$^3$ was used for all values of $\theta$, giving approximately 118 bacteria in each aggregate.

\section{Generating configurations of competing ``single cells" in 3D}
To surround the 3D aggregates with neighbours, we seeded the surface surrounding the aggregate with single cells. The number of these competing cells per square $\mu$m gave the density of cells on the surface.  Cells that lay in the region defined by the radius of the spherical cap, $a$, were removed from this configuration. 
The aggregate configuration and the surrounding cell configuration were then combined to produce the initial state of the system. 

\section{Multiple simulation runs in 2D and 3D}\label{sec:repeats}
For our 2D simulations, four different aggregate configurations were generated for each value of $\theta$, and for each of these configurations, 5 simulations of 480 hours were performed using a different initial distribution of surrounding cells. This gave rise to a total of 20 simulations for each value of $\theta$. 

In 3D, only the spread aggregate, $\theta = 5^o$, and the rounded aggregate, $\theta = 180^o$, were simulated. 
Three different initial configurations were generated for each value of $\theta$, and for each of these configurations 3 simulations of 72 hours were performed, each using different initial distributions of surrounding single cells. This gave rise to a total of 9 simulations for each value of $\theta$.

Note that, in both the 2D and 3D simulations, the competitor cells on the surface and the aggregated cells are identical in their growth parameters (Table 1, main manuscript).

\section{Exploration of Biofilm Structure at Low and High Nutrient Concentration Without Aggregates} \label{sec:conc}
\begin{figure*}[t!]
\begin{center}
\includegraphics[scale=0.4]{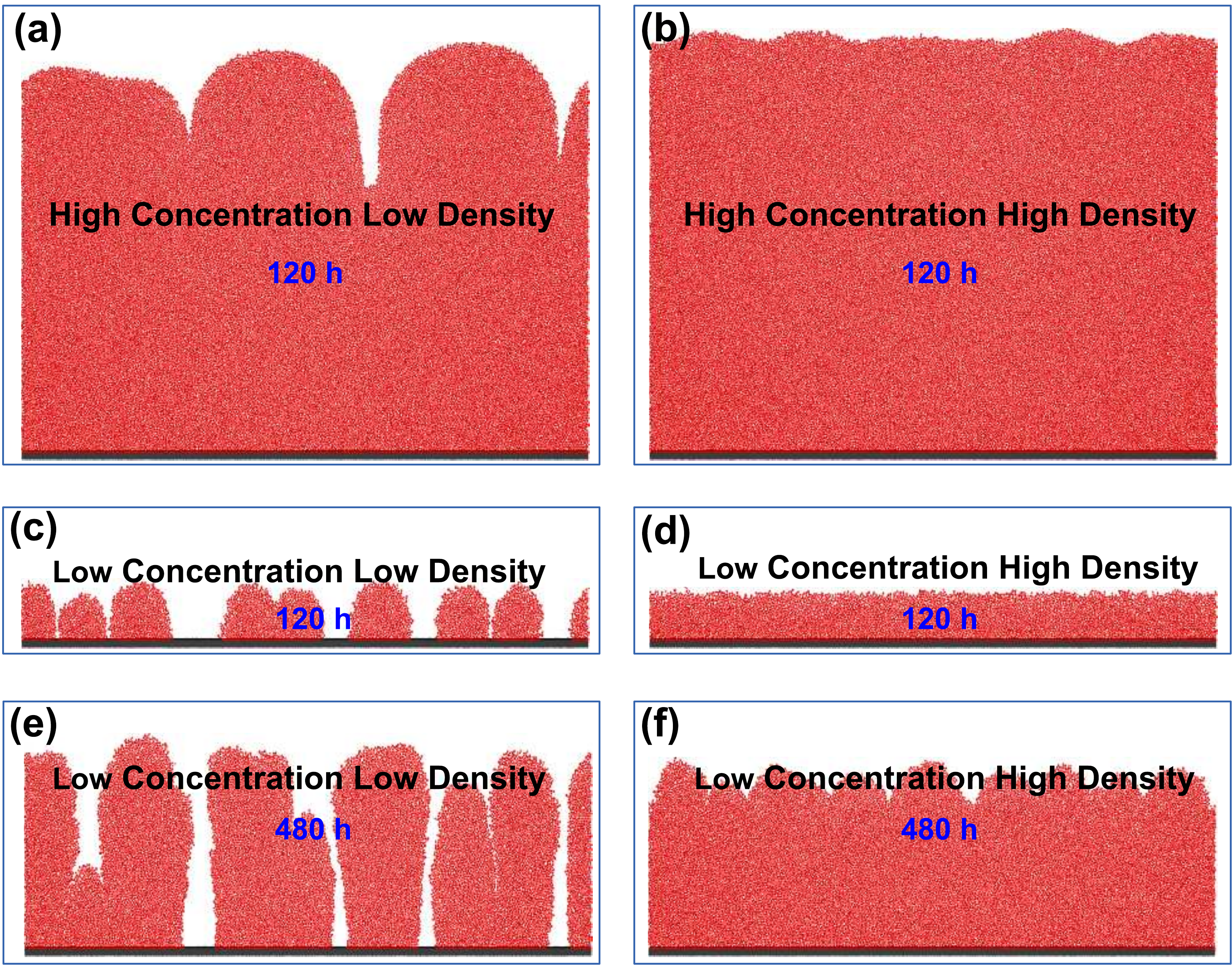}  \\
\caption[]{Nutrient concentration and initial density of seeder cells affect biofilm morphology in the absence of initial aggregates: (a) Nutrient concentration $= 5.4 \times 10^{-2}$g L$^{-1}$, Density of seeder cells $= 0.01$ cell $\mu$m$^{-1}$. (b) Nutrient concentration $= 5.4 \times 10^{-2}$g L$^{-1}$, Density of seeder cells $= 0.5$ cell $\mu$m$^{-1}$. (c) Nutrient concentration $= 5.4 \times 10^{-3}$g L$^{-1}$, Density of seeder cells $= 0.01$ cell $\mu$m$^{-1}$. (d) Nutrient concentration $= 5.4 \times 10^{-3}$g L$^{-1}$, Density of seeder cells $= 0.5$ cell $\mu$m$^{-1}$. (e) Nutrient concentration $= 5.4 \times 10^{-3}$g L$^{-1}$, Density of seeder cells $= 0.01$ cell $\mu$m$^{-1}$, 480 h. (f) Nutrient concentration $= 5.4 \times 10^{-3}$g L$^{-1}$, Density of seeder cells $= 0.5$ cell $\mu$m$^{-1}$, 480h.}
\label{fig:conc_regime}
\end{center}
\end{figure*}
To investigate the effect of nutrient concentration in general in our simulations, we simulated 2D biofilms in the absence of initial aggregates, at several different nutrient concentrations and various densities of cells, which were initially located at random across the surface.
Figures \ref{fig:conc_regime} (a) and (b) show the resulting biofilms formed after 120 h at a nutrient concentration of $5.4 \times 10^{-2}$g L$^{-1}$ for seeder cell densities of 0.01 cell $\mu$m$^{-1}$ and 0.5 cell $\mu$m$^{-1}$. 
Clearly, the biofilms formed at this nutrient concentration are significantly thicker than those formed at lower nutrient concentration (Figures \ref{fig:conc_regime}(c) to (f)).
Comparing (a) and (b), we see that a low density of seeder cells leads to a less uniform biofilm morphology with finger-like projections. 

After 120 h, at low nutrient concentrations ($5.4 \times 10^{-3}$g L$^{-1}$), the biofilms remain very thin (Figures \ref{fig:conc_regime}(c) and (d)). At this low concentration, the initial number of seeder cells has a marked effect on the final structure of the biofilm. A high density of seeder cells yields a biofilm with a uniform morphology whereas a low density of seeder cells yields a biofilm with a spatial structure that is clearly dependent on the initial configuration of the seeder cells. Figures \ref{fig:conc_regime}(e) and (f) show that this effect persists at long times. 

From Figures \ref{fig:conc_regime}(a) and (e), it is clear that both longer times and/or higher nutrient concentration lead to larger biofilms. Selecting a bulk nutrient concentration and timescale to investigate the shape dependent fitness of aggregated clusters of bacteria is therefore non-trivial.   
In our simulations, we used the lower nutrient concentration of 
$5.4 \times 10^{-3}$g L$^{-1}$ (see Table 1 in the main manuscript), comparable with previous biofilm simulations \cite{Xavier2005, Xavier2007, Lardon2011}. We also chose to run the simulations for 480 h in order to explore the long term growth dynamics, and to generate populations that were large enough to obtain good statistics.  Timescales of 480 h have also been used in previous simulation studies \cite{Xavier2005, Xavier2007}. After 480 h growth using a bulk concentration of $5.4 \times 10^{-3}$g L$^{-1}$, our simulations produce biofilms that are approximately 200-300 $\mu$m in height (Figures \ref{fig:conc_regime}(e) and (f)). 

\section{Exploration of Aggregate Fate at Different Times}\label{sec:time}
\begin{figure}[tb]
\begin{center}
\includegraphics[scale=1.2]{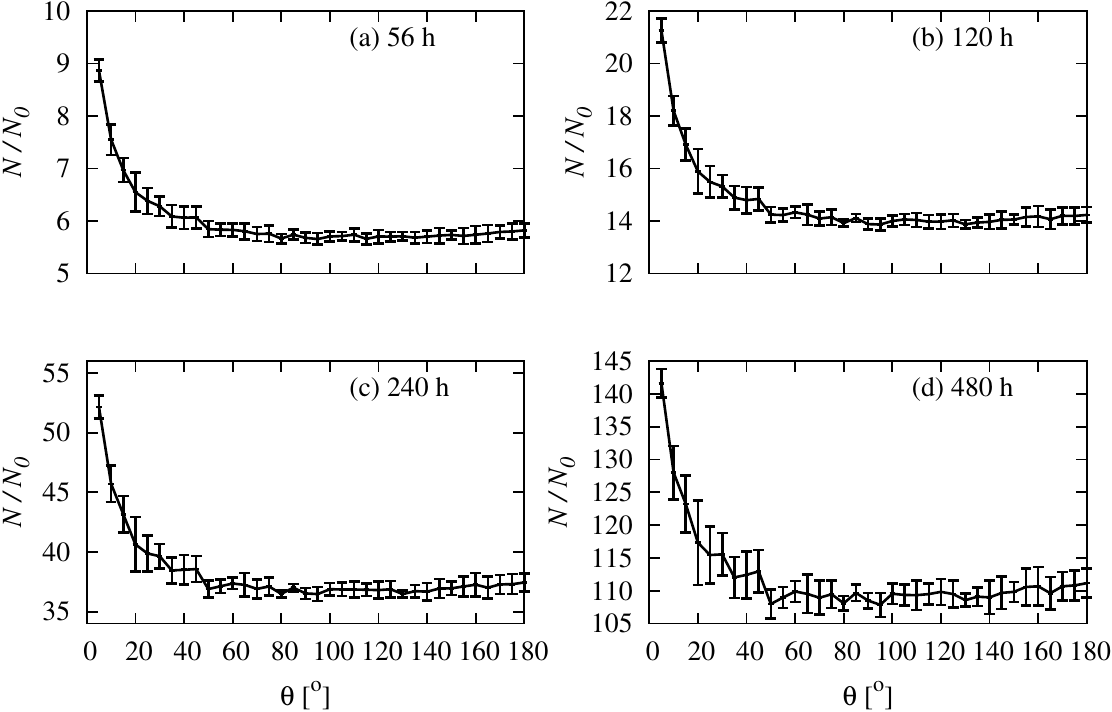} 
\caption[]{Without competing cells on the surface, the functional form of the $N/N_0$ curves remains relatively constant over time. Note, however that the scale does change because more cells are produced as time increases: (a) 56 h. (b) 120 h. (c) 240 h. (d) 480 h. Error bars represent the standard deviation from 20 simulations.}
\label{fig:fit_time_no_comp}
\end{center}
\end{figure}

In the main manuscript, we computed $N/N_0$, as a function of the aggregate surface angle, $\theta$, after 480 h. Here, we show that qualitatively similar results are obtained regardless of the time at which fitness is measured.  
Figures \ref{fig:fit_time_no_comp} shows $N/N_0$ as function of $\theta$ in the absence of competing cells (no red cells in Figure 1(a) of the main manuscript) at 4 different times. Evidently, the functional form of the curves does not change significantly with time. 


\begin{figure}[tb]
\begin{center}
\includegraphics[scale=1.2]{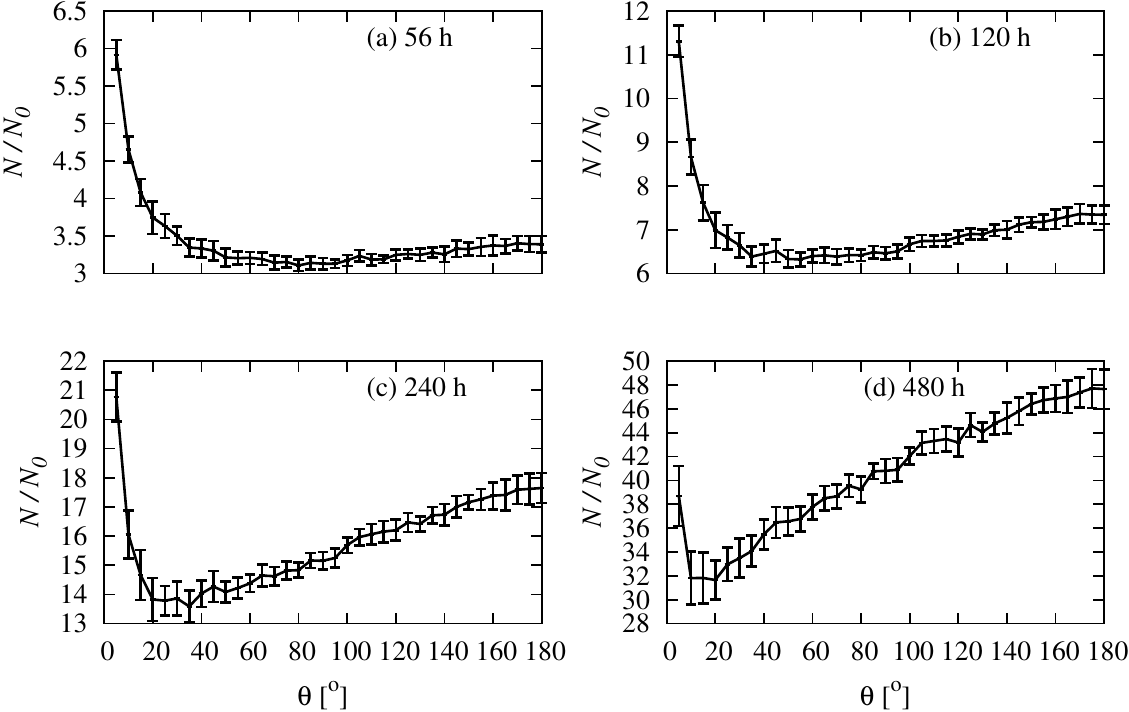} 
\caption[]{With competing cells on the surface at a density of 0.5 cell $\mu$m$^{-1}$, the functional form of the $N/N_0$ curve changes over time: (a) 56 h. (b) 120 h. (c) 240 h. (d) 480 h. Error bars represent the standard deviation from 20 simulations.}
\label{fig:fit_time_comp}
\end{center}
\end{figure}

In the highly competitive regime where we include competitor cells (red cells in Figure 1 of main manuscript) at a density of 0.5 cells $\mu$m$^{-1}$, the situation is more complex. Here, the functional form of the curves changes during the course of the simulation (Figure \ref{fig:fit_time_comp}). For early times, the curves are similar to the case without competition, suggesting that shape-dependent nutrient access is the still dominant factor in the early stages. As time increases, the rounded aggregates become increasingly ``fitter" relative to the spread aggregates, until at 480 h, all aggregates with a surface aggregate angle greater than 75$^o$ are fitter than the spread aggregate. From this we can conclude that the initial height advantage of the rounded aggregate only becomes important at longer times. 

\section{Difference in Fate Between Spread and Rounded Aggregates at Higher Nutrient Concentration} \label{sec:higher_conc}

\begin{figure}[t!]
\begin{center}
\includegraphics[scale=1.8]{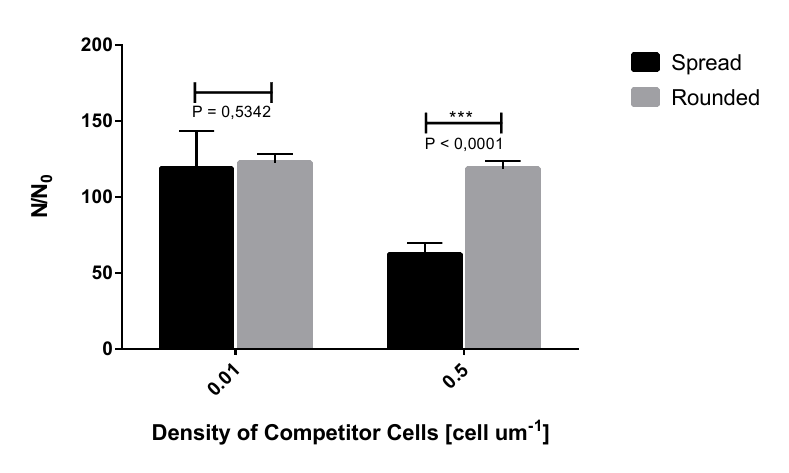}  \\
\caption[]{At higher nutrient concentration, $1\times10^{-2}$gL$^{-1}$, rounded aggregates are more favourable for growth. At low density of seeder cells (0.01 cell $\mu$m$^{-1}$) there is no significant difference in $N/N_0$ between the spread and rounded aggregates. At high density (0.5 cell $\mu$m$^{-1}$), the rounded aggregate is again produces more progeny per initial cells relative to that of the spread. Note that these data were generated from simulations representing 480 h of biofilm growth.}
\label{fig:higher_conc}
\end{center}
\end{figure}
Here, we investigate the fate of the spread and round aggregate at a higher nutrient concentration, $1\times10^{-2}$ gL$^{-1}$. Figure \ref{fig:higher_conc} shows that our conclusions remain valid at this nutrient concentration: at low density of seeder cells there is no difference in fate between the spread and rounded aggregates, whereas at higher density of initial neighbours, the rounded aggregate is favoured over its initially more spread counterpart. Note that these simulations were again run for 480 h.

\section{The Effect of Aggregate Shape on the Growth of the Active Layer} \label{sec:act_layer}
In the main manuscript we discussed cell growth rate heterogeneity within the biofilm and the presence of a well-defined active layer in the growing aggregate. The depth of the active later has been shown to be very important in determining biofilm structure and the degree of segregation amongst cell groups within the community \cite{Nadell2010}. It has also been discussed in extensive mathematical detail by Dockery and Klapper \cite{Klapper2001}. Here we look at the growth of the active layer.

In the analysis that follows, we simply define the active layer as the region of the growing aggregate in which all cells grow with a growth rate greater than 1 fg h$^{-1}$ (see Figure 5 of main manuscript), and we assess the number of cells in this layer. This choice is arbitrary; what is important is the qualitative difference in behaviour between the aggregate shapes.

\begin{figure}[t!]
\begin{center}
\includegraphics[scale=0.8] {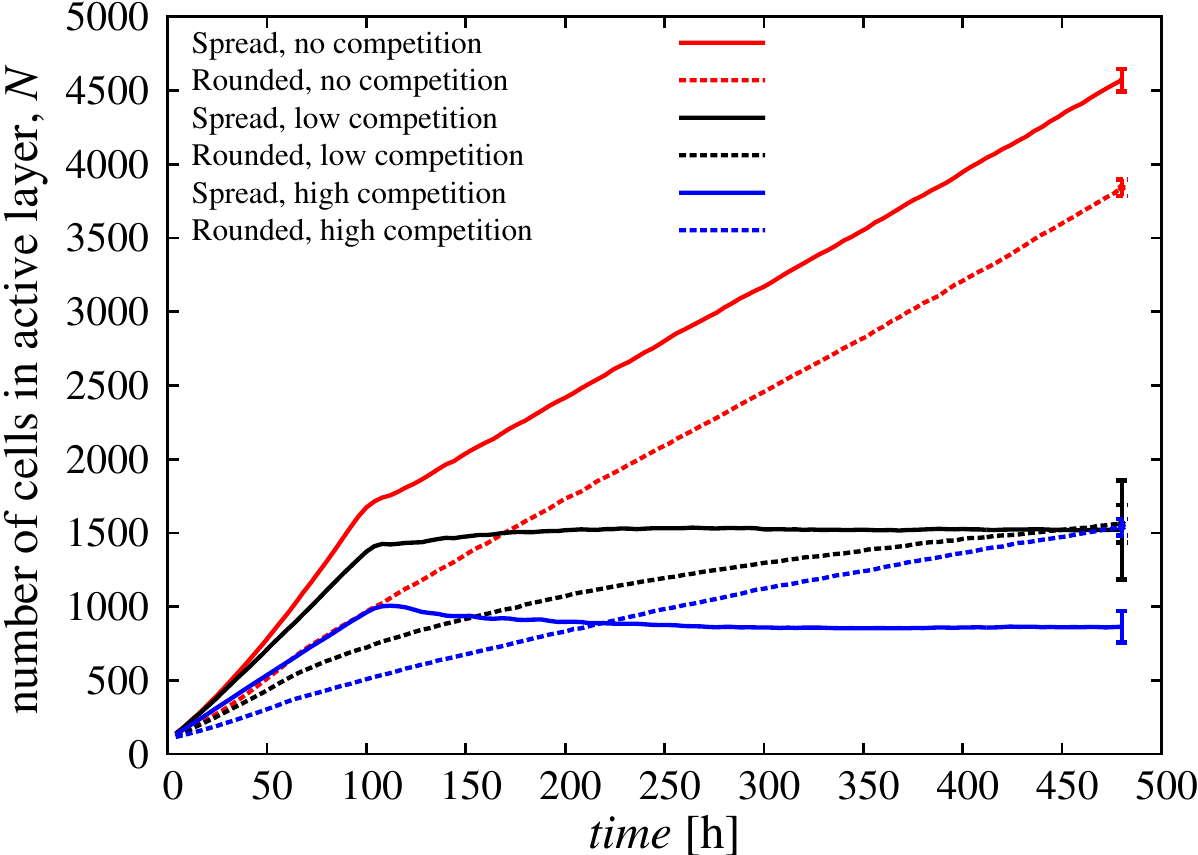}\\
\caption[]{Dynamics of the growing layer determine aggregate growth behaviour. The number of growing cells in the spread and rounded aggregates at various competitive regimes. For clarity the error bars, representing the standard deviations, are only shown for the final data points. The standard deviations at these points are maximal.}
\label{fig:act_layer}
\end{center}
\end{figure}

Figure \ref{fig:act_layer}(a) shows the number of active cells within the spread ($\theta = 5^o$) and rounded ($\theta = 180^o$)  aggregates at varying degrees of competition; dictated by the density of surrounding unaggregated cells on the surface. In the absence of competition, the number of active cells within the spread and rounded aggregates increases with time. Initially the number of active cells in the spread aggregate grows faster than that for its rounder counterpart, however for $t >100$ h, the number of active cells in both aggregates increases with similar rates (slope of red curves). This linear increase determines the super-linear increase in the corresponding aggregate populations in Figure 1(b) of the main manuscript: a linear increase in the number of active cells translates to a super-linear increase in the total aggregate population.

In all competitive regimes, the number of cells in the growing layer of the spread aggregates becomes constant after $\sim$100 h. A constant population of active cells in the growing layer translates to linear growth behaviour of the total aggregate population (compare solid blue curve with the dashed red curve in Figure 1(b) of the main manuscript). For the rounded aggregates, this transition to the constant regime does not occur within the timescales of our simulations, and the layer continues to grow monotonically. 

Comparing the spread and rounded aggregates (solid curves with dashed curves), with competition, we see that growth of the active layer always increases faster for the spread aggregate before levelling off after 100 h. However, later in the simulation, the number of active cells in the rounded aggregates becomes greater. The time at which this crossover occurs increases with increasing levels of competition. The increasing growth of the active layer in the rounded aggregate at high competition, $\rho = 0.5$ cell $\mu$m$^{-1}$, explains why its number of progeny becomes larger than that of the spread aggregate in Figure 1(b), and why its structures tend to fan outwards at the top (Figure 4(d)).

\section{Aggregate Simulations in 3D}\label{sec:3D_results}
\begin{figure}[t!]
\begin{center}
\includegraphics[scale=0.28]{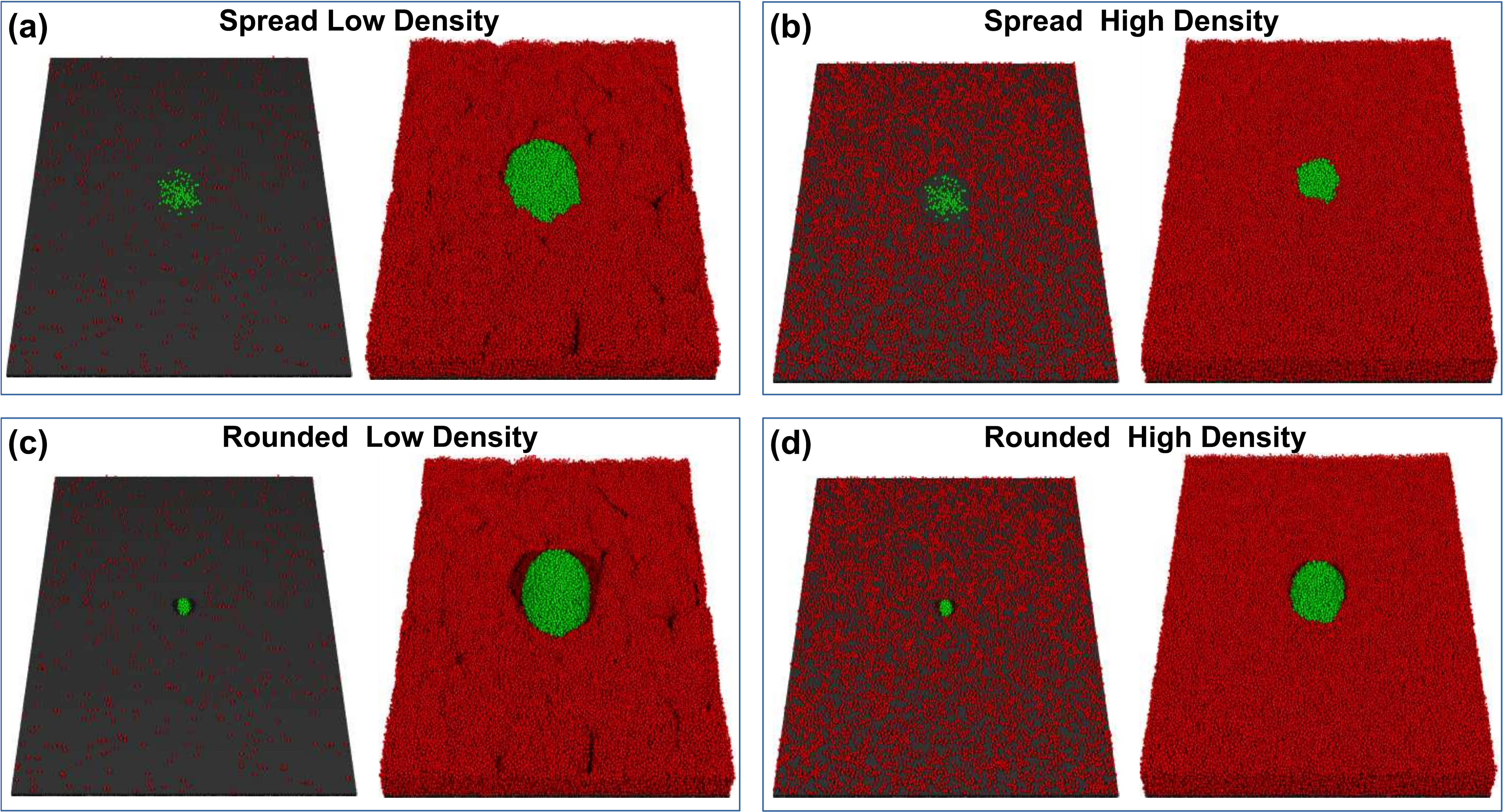}  \\
\caption[]{3D simulation snapshots showing initial and final configurations of the spread and rounded aggregates at low and high density of surrounding cells. (a) Spread aggregate, Density of seeder cells = 0.002 cell $\mu$m$^{-2}$. (b) Spread aggregate, Density of seeder cells = 0.03 cell $\mu$m$^{-2}$. (c) Rounded aggregate, Density of seeder cells = 0.002 cell $\mu$m$^{-2}$. (d) Rounded aggregate, Density of seeder cells = 0.03 cell $\mu$m$^{-2}$.}
\label{fig:3D_snaps}
\end{center}
\end{figure}

The results presented in the main manuscript are for 2D simulations, for reasons of computational feasibility.
To validate our results further, we also ran some simulations in 3D. Figure \ref{fig:3D_snaps} shows simulation snapshots after 72 h for the spread and rounded aggregates in the presence of low and high densities of competitor cells. As in our 2D simulations (Figures 3(b) and (d) of the main manuscript), the rounded aggregate persists as a dominant structural feature at both low and high densities of competitor cells. The spread aggregate on the other hand becomes swamped by the surrounding competitors and after 72 h is structurally indistinguishable from the rest of the biofilm (as in our 2D simulations, Figures 3(b) and (d) of the main manuscript). 

\begin{figure}[t!]
\begin{center}
\includegraphics[scale=0.85]{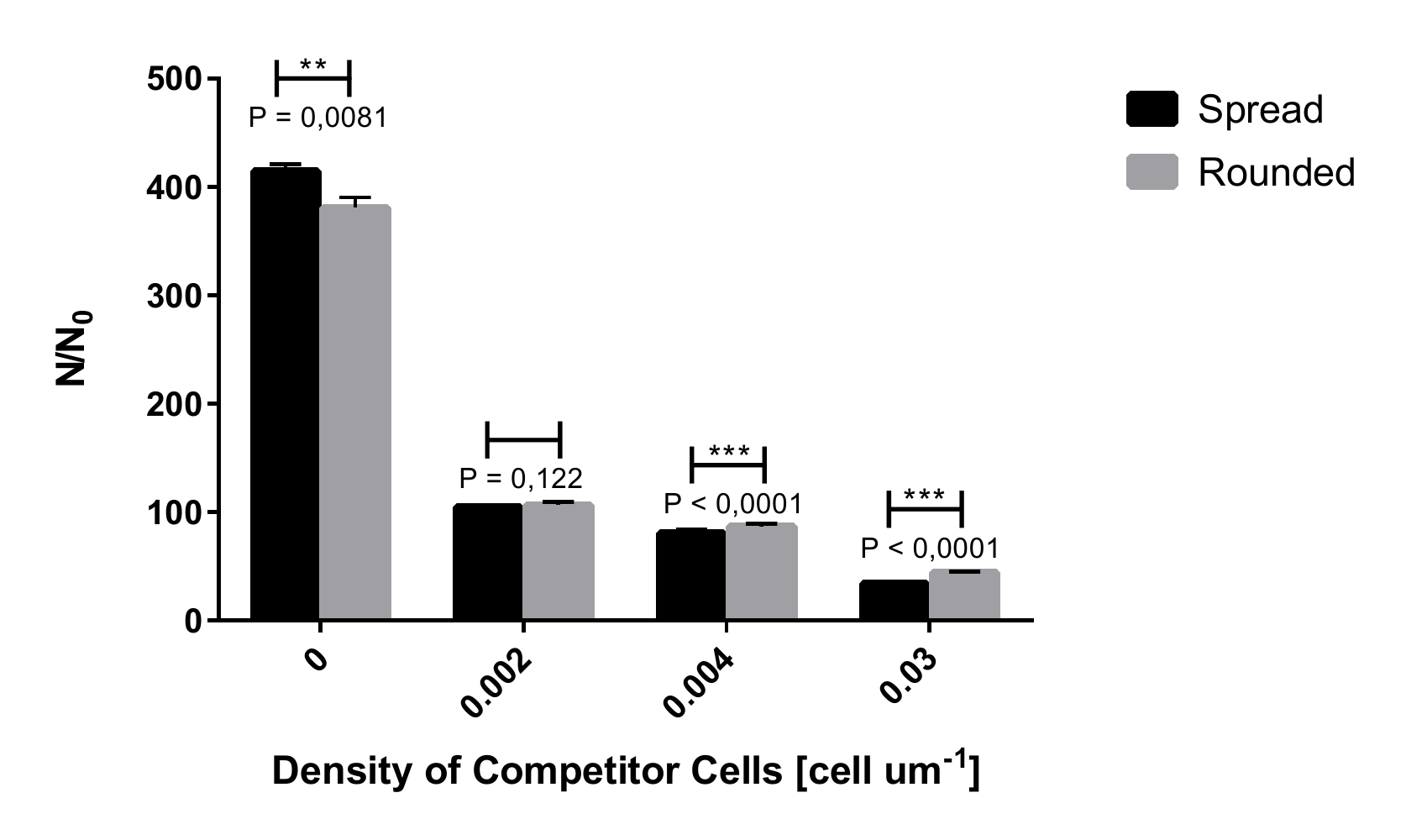}  
\caption[]{$N/N_0$ of the rounded aggregate with respect to the spread aggregate increases with the density of neighbouring cells on the surface. In 3D, the trend is similar to that observed in 2D. The spread aggregate is significantly favoured when there is no competition on the surface however with increasing density the rounded aggregate is more favourable for growth.}
\label{fig:3D_delta_f}
\end{center}
\end{figure}

$N/N_0$ for the spread and rounded aggregates are plotted in Figure \ref{fig:3D_delta_f} and is analogous to the data shown in Figures 8 of the main manuscript for the 2D simulations. As in the 2D case, the number of progeny per initial cell of the aggregates decreases with increasing density of surrounding cells. In the absence of competition, the spread aggregate again produces significantly more progeny than the rounded aggregate, however as the competition on the surface is increased, the number of progeny produced by cells in the rounded aggregate increases relative to that of the spread.

 
